\documentclass[twocolumn]{aastex701}
\usepackage{amsmath}
\usepackage{rotating}

\newcommand\msun{\ensuremath{\text{M}_\odot}}

\newcommand\kms{\ensuremath{\text{km}\,\text{s}^{-1}}}

\newcommand{\rbin}{\ensuremath{r_{\text{bin}}}}

\newcommand{\tff}{\ensuremath{t_{\text{ff}}}}
\newcommand{\tcross}{\ensuremath{t_{\text{cross}}}}
\newcommand{\texp}{\ensuremath{t_{\text{exp}}}}
\newcommand{\mpl}{\ensuremath{M_{\text{pl}}}}
\newcommand{\rpl}{\ensuremath{R_{\text{pl}}}}
\newcommand{\rhalf}{\ensuremath{R_{\text{h}}}}
\newcommand{\drdt}{\ensuremath{dR_{\text{h}}/dt}}
\newcommand{\vout}{\ensuremath{v_{\text{R}}}}
\newcommand{\sigmavr}{\ensuremath{\sigma(v_r)}}

\newcommand{\sigmamax}{\ensuremath{\sigma_{\rm *,max}}}
\newcommand{\fret}{\ensuremath{F_{\text{ret}}}}
\newcommand{\fretmax}{\ensuremath{F_{\text{ret,max}}}}

\newcommand{\starforgefiducialfmax}[1]{%
\starforgefiducial\_%
\ifnum#1=0
\texttt{match}%
\else
\texttt{fmax#1}%
\fi}

\newcommand{\torchclaudefmax}[1]{%
\torchclaude\_%
\ifnum#1=0
\texttt{match}%
\else
\texttt{fmax#1}%
\fi}

\newcommand{\starforgefiducial}{\texttt{starforge\_fiducial}}
\newcommand{\torchclaude}{\texttt{torch\_M1}}

\usepackage[dvipsnames]{xcolor}

\begin{document}

\title{Dynamical Cluster Assembly Framework (D-CAF): \\The Link Between Star Cluster Formation and Expansion Rates
}

\author[orcid=0000-0002-5851-2602,sname='Farias']{Juan P. Farias}
\affiliation{Department of Physics and Astronomy, McMaster University, 1280 Main Street W, Hamilton, ON L8S 4L8, Canada}
\email[show]{fariasoj@mcmaster.ca}  

\author[orcid=0000-0003-3551-5090, sname='Sills']{Alison Sills} 
\affiliation{Department of Physics and Astronomy, McMaster University, 1280 Main Street W, Hamilton, ON L8S 4L8, Canada}
\email{}

\begin{abstract}
We introduce the Dynamical Cluster Assembly Framework (D-CAF), an AMUSE-based
framework designed to connect embedded star formation histories to the dynamical
evolution of young stellar systems. We model star formation through the gradual
formation of stars inside an evolving background potential, where the global gas
evolution is extracted from realistic magneto-hydrodynamical (MHD)
simulations. In this first work, we focus on the global evolution of the natal
gas and its dynamical imprint on the stellar population. Across all explored
MHD setups, we find that the gas continues to collapse while stars are forming,
increasing both the central concentration and velocity scale of the embedded
stellar population before gas expulsion. Using a controlled grid of direct
$N$-body simulations, we show that this embedded evolution strongly regulates
both the survival and later expansion of young stellar systems. In particular,
gas contraction shortens the stellar crossing time prior to gas expulsion,
making the same gas-removal timescale effectively more adiabatic for the stars.
We find that the present-day expansion of stellar associations still preserves
information about the embedded dynamical state reached during formation. The
expansion rate is limited by the velocity scale reached before gas expulsion,
while the efficiency with which this velocity field is transformed into
expansion depends on the gas-expulsion timescale. Finally, we show that some
commonly used expansion diagnostics can directly trace the physical expansion
rate of young stellar systems when full kinematic information is available,
opening the possibility of using stellar kinematics to constrain the dynamical
conditions of embedded star formation.
\end{abstract}

\section{Introduction}

Most stars form in clustered environments, at densities far above the field, yet
a large fraction of young stars are dispersed within only a few Myr
\citep{lada2003,bressert2010,pokhrel2020}. Star formation takes place inside
molecular clouds with intrinsically complex dynamics. The gas is turbulent and
magnetised, strongly structured over a wide range of scales, and rapidly
evolving under its own gravity and stellar feedback
\citep{maclow2004,elmegreen2007}. The interplay between these processes, and
their relative importance across different environments, is therefore expected
to play a crucial role in setting the early survival rate and kinematic properties 
of star clusters.

Given this complexity, realistic magneto-hydrodynamic (MHD) simulations have
become essential for understanding how stars assemble in such environments and
how their feedback modifies the surrounding gas during formation
\citep{grudic2019,federrath2012,dale2013a}. However, this realism comes at a
high computational cost. It is currently not possible to run sufficiently large suites of 
state-of-the-art simulations to (i) explore large ensembles across initial
conditions and star formation prescriptions, (ii) follow star clusters for
significant periods after their formation -- when the dynamical consequences of
the initial conditions become recognizable -- and (iii) disentangle which
outcomes are a direct consequence of the star formation process and which arise
from the intrinsic stochasticity of stellar dynamics. Many important phenomena
are intrinsically rare, such as the production of runaway stars, the detailed
kinematics of O and B star populations, or the emergence and survival of
sub-clusters in different turbulent realisations
\citep{fujii2011,smith2011,parker2014,farias2015}. Robustly characterising such
effects therefore requires large sets of statistically equivalent simulations,
which remains prohibitively expensive with current computational resources.

At the same time, many of the key observables used to link star formation to
present-day stellar populations are fundamentally dynamical in nature. These
include the bound fraction after gas dispersal
\citep{goodwin2006,baumgardt2007}, the development of mass segregation and the
survival and interacting effects of binaries and higher-order multiples
\citep{moeckel2010,parker2011,pavlik2020}. Addressing these questions requires
long integrations and controlled ensembles, conditions under which direct
$N$-body methods excel, provided they are supplied with a physically grounded,
time-dependent description of the natal environment in which the stars formed.

This is why simplified dynamical models remain essential. They allow rapid,
controlled experiments and large ensembles, making it possible to isolate the
impact of specific physical ingredients without re-running full MHD each time.
The key is to make the simplifications meaningful, retaining the relevant
consequences of the underlying physical processes reducing computational cost
and unnecessary degrees of freedom, while retaining the overall environment that
most strongly conditions the stellar dynamics.

In this paper we introduce the Dynamical Cluster Assembly Framework (D-CAF),
implemented within AMUSE \citep{portegieszwart2013,pelupessy2013}, to model star
cluster formation as the gradual assembly of stars within an evolving external
potential whose global evolution is motivated by realistic MHD
simulations.  Our goal is not to reproduce the full hydrodynamics, but to retain
the time-dependent environment that matters most for stellar dynamics, enabling
systematic and long-term studies that are impractical with full MHD alone.

In this first work we focus on the global evolution of the natal gas in suites
of MHD simulations, asking what aspects of that evolution are robust across
setups, and what their dynamical consequences are for cluster survival under gas
dispersal.

\section{Methods}

The goal of this work is to identify the large-scale, time-dependent gas
environment in which stars form, and to study how this environment affects the
subsequent dynamical evolution of the stellar system. To do this, we analyse
MHD star formation simulations, extract the time evolution of the natal gas,
and use this information to build dedicated $N$-body models with
time-dependent background potentials.

For this purpose, we developed the Dynamical Cluster Assembly Framework (D-CAF),
which separates the problem into two parts. First, we measure the global gas
evolution from the MHD simulations. Second, we evolve the forming stellar
population inside that background using direct $N$-body simulations. This lets
us keep the main large-scale effect of the gas while avoiding the computational cost of
following the full hydrodynamics. With this approach, we can isolate the effect
of the underlying star formation prescription on the long-term stellar
dynamics.

\subsection{Framework}\label{sec:dcaf}

We built D-CAF inside the AMUSE framework, which we use to couple the different
parts of the calculation and to handle the data output. The framework has three
main components: i) the $N$-body integrator, which evolves the stellar orbits,
ii) the user-provided star formation scheduler (hereafter framework), which
specifies when and how stars are formed, and iii) the time-dependent background
potential, which describes the large-scale evolution of the natal gas. These
components are modular and can be easily replaced in order to adapt to new
setups.

The system can in principle be extended to include additional modules, such as
stellar evolution, although we do not include these in this first work. Our goal
here is not to build a full-physics simulation, but to isolate the stellar
dynamical response to a time-evolving gas background under a controlled star
formation prescription. What is considered sufficiently realistic is therefore
set by the problem being studied.

We couple the $N$-body integrator and the background gas using BRIDGE
\citep{fujii2007}, as implemented in AMUSE. In this scheme, the stellar system
feels the gravitational force of the background potential at fixed coupling
intervals. We represent the gas with an analytical model whose evolution can be
given either by an explicit functional form or by parameters tabulated in time,
in which case we interpolate the potential parameters between outputs.

For the stellar dynamics we use \textsc{PeTar} \citep{wang2020}, a collisional
$N$-body code that combines a fourth-order Hermite scheme for close interactions
with a Barnes \& Hut tree algorithm \citep{barnes1986} for long-range forces.
\textsc{PeTar} also implements a regularization scheme for close binaries within
a user-defined radius $\rbin$ \citep[see][for details]{wang2020}.

We choose the BRIDGE coupling timestep to be no smaller than the internal
timestep of the $N$-body integrator, so that the stellar self-gravity and the
external potential follow a consistent temporal ordering. In practice, we set
the BRIDGE timestep to an integer multiple of the $N$-body timestep, with both
written as exact powers of two. This choice simplifies the synchronization
between the different parts of the calculation and avoids numerical artefacts
from incommensurate timesteps.

\subsection{Target MHD simulations}\label{sec:mhd}

The first step in our problem is to identify the evolving gas environment in
which the stars form. In practice, this means following how the gas
distribution changes with time in the star-forming region. To do this, we
analyse MHD simulations of molecular clouds that include self-gravity and
stellar feedback. From these simulations, we extract the time evolution of the
gas mass distribution, central concentration, and characteristic spatial scales
of the natal environment.

We use star cluster formation simulations produced with two independent
frameworks that follow different approaches to star formation: the Star
FORmation in Gaseous Environments (STARFORGE) simulations
\citep{grudic2021a,guszejnov2023} and the TORCH framework \citep{wall2019,wall2020}.
STARFORGE models star formation by resolving individual stars as sink particles
that accrete gas self-consistently and includes a broad set of stellar feedback
processes. However, STARFORGE uses a non-specialized $N$-body treatment that
does not fully resolve very close encounters and hard binaries. TORCH instead
uses a modular framework in which stars form from a predefined list sampled
from an initial mass function and are coupled to the gas through sink
particles. TORCH includes a more limited set of feedback processes, but it
adopts observationally motivated binary populations and uses a dedicated
$N$-body solver that resolves hard binaries and close encounters more
accurately. Using both frameworks lets us test which trends are robust across
different star formation prescriptions and which depend more strongly on the
details of the stellar dynamics. Below, we briefly describe each framework and
the specific simulations used in this work.

\subsubsection{STARFORGE simulations}

The STARFORGE framework is implemented in the MHD code GIZMO
\citep{hopkins2015,grudic2021}. It uses the Lagrangian meshless finite-mass
(MFM) method to solve the MHD equations under the ideal MHD approximation
\citep{hopkins2015,hopkins2016}, together with an improved Barnes \& Hut
algorithm \citep{springel2005} to compute self-gravity. These simulations
resolve the formation of individual stars with sink particles and include the
main stellar feedback channels, such as radiation pressure, photoionization,
stellar winds, and supernovae
\citep[e.g.][]{grudic2018,grudic2021a,guszejnov2021}. Gravitational interactions
between sink particles are solved with a fourth-order Hermite integrator.
However, unlike PeTar, STARFORGE does not include regularization for close
encounters and binaries. It therefore uses a softening radius of 80\,AU to avoid
prohibitively small timesteps.

We use the fiducial simulation from \cite{guszejnov2022a}, their \texttt{M2e4}
model which correspond to initially spherical turbulent molecular clouds with
mass $2\times10^4\,\msun$ and radii of 10\,pc with uniform density. We refer to
this run as the \starforgefiducial~ model.

\subsubsection{TORCH simulations}

TORCH is a modular framework developed in
AMUSE that couples the gas and stellar dynamics through BRIDGE
\citep{fujii2007}. It evolves the gas with the adaptive mesh refinement MHD
solver FLASH \citep{fryxell2000}, while it follows stellar dynamics with
the direct $N$-body solver PeTar. TORCH includes radiative feedback self-consistently
through the radiation hydrodynamics module implemented in FLASH, as well as
stellar winds and supernova feedback, but it does not include protostellar
outflows in the simulations used here.

We focus on the \texttt{M1} simulation presented in
\citet{cournoyer-cloutier2024a}, hereafter \torchclaude. This simulation
follows a $2\times10^4\,\msun$ turbulent spherical cloud with a Gaussian
initial density profile. The cloud is initialized with supersonic turbulence
and then allowed to collapse under self-gravity until feedback regulates
further star formation.

As in STARFORGE, TORCH forms stars through sink particles. The main difference
is that these sink particles do not represent individual stars. Instead, TORCH
forms stars from a predefined list sampled from a \citet{Kroupa2002} initial
mass function, including the observationally motivated binary prescription
introduced by \citet{cournoyer-cloutier2020}.

\begin{figure*}
    \centering
    \includegraphics[width=\textwidth]{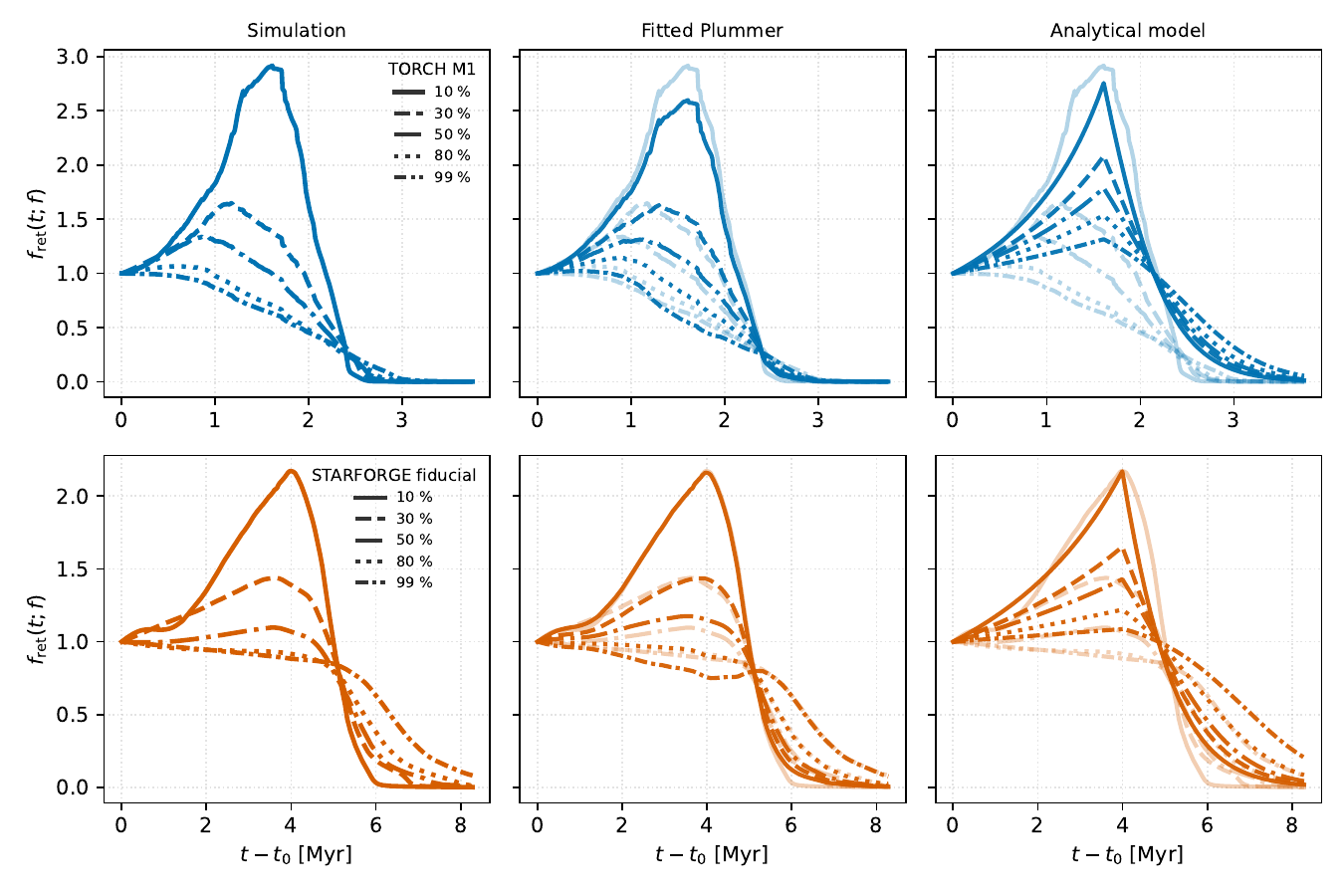}
    \caption{Evolution of the retained gas fraction, $\fret(t; f)$, measured
    within fixed initial gas Lagrangian radii for the \torchclaude\ (top row)
    and \starforgefiducial\ (bottom row) simulations. The left panels show the
    values measured directly from the MHD simulations. The middle panels show
    the corresponding $\fret(t)$ reconstructed from the fitted Plummer profiles
    at each snapshot. The right panels show the analytic model used to
    parametrize the gas evolution. In the middle and right panels, the fainter
    lines show the MHD simulation results for comparison.}
    \label{fig:fret_evol}
\end{figure*}

\subsection{Extracting gas evolution from MHD simulations}
\label{sec:gas_extraction}

To build the time-dependent background gas model used in D-CAF, we extract the
large-scale gas evolution from the MHD simulations described in
\S\ref{sec:mhd}. Our goal is not to reproduce the full gas dynamics, but to
capture the time-dependent gravitational environment in which stars form and
evolve.

Throughout this work, we define $t_0$ as the time at which the first star
forms. To characterize the part of the gas evolution that is dynamically
relevant for the stars, we measure the fraction of gas retained within a fixed
set of initial gas Lagrangian radii, $\fret$, as a function of time. This
quantity gives a simple measure of how the gas redistributes during the
simulations. The left panels of Figure~\ref{fig:fret_evol} show the evolution
of $\fret$ for the MHD simulations considered in this work. The behaviour is
broadly consistent with the two-phase picture assumed in classical studies: an
embedded phase, during which the gas collapses and becomes more centrally
concentrated while stars are forming, followed by a gas-expulsion phase driven
by stellar feedback.

For the D-CAF models, we use two different descriptions of this gas evolution.
The first uses the measured gas evolution directly through tabulated background
potentials, while the second approximates the same evolution using an analytic
parameterization. These two approaches are shown in the middle and right panels
of Figure~\ref{fig:fret_evol}, and are described below.

To connect this global evolution to the spatial structure of the gas, we also
compute radial gas density profiles at each available snapshot, centred on the
simulation origin. We average the density in equal-volume shells, which filters
out small-scale substructure and anisotropy while retaining the overall radial
distribution of the gas. We use a fixed region of the simulation domain, chosen
to contain 90\% of the stellar mass at all times.

\begin{figure*}
    \centering
    \includegraphics[width=0.49\textwidth]{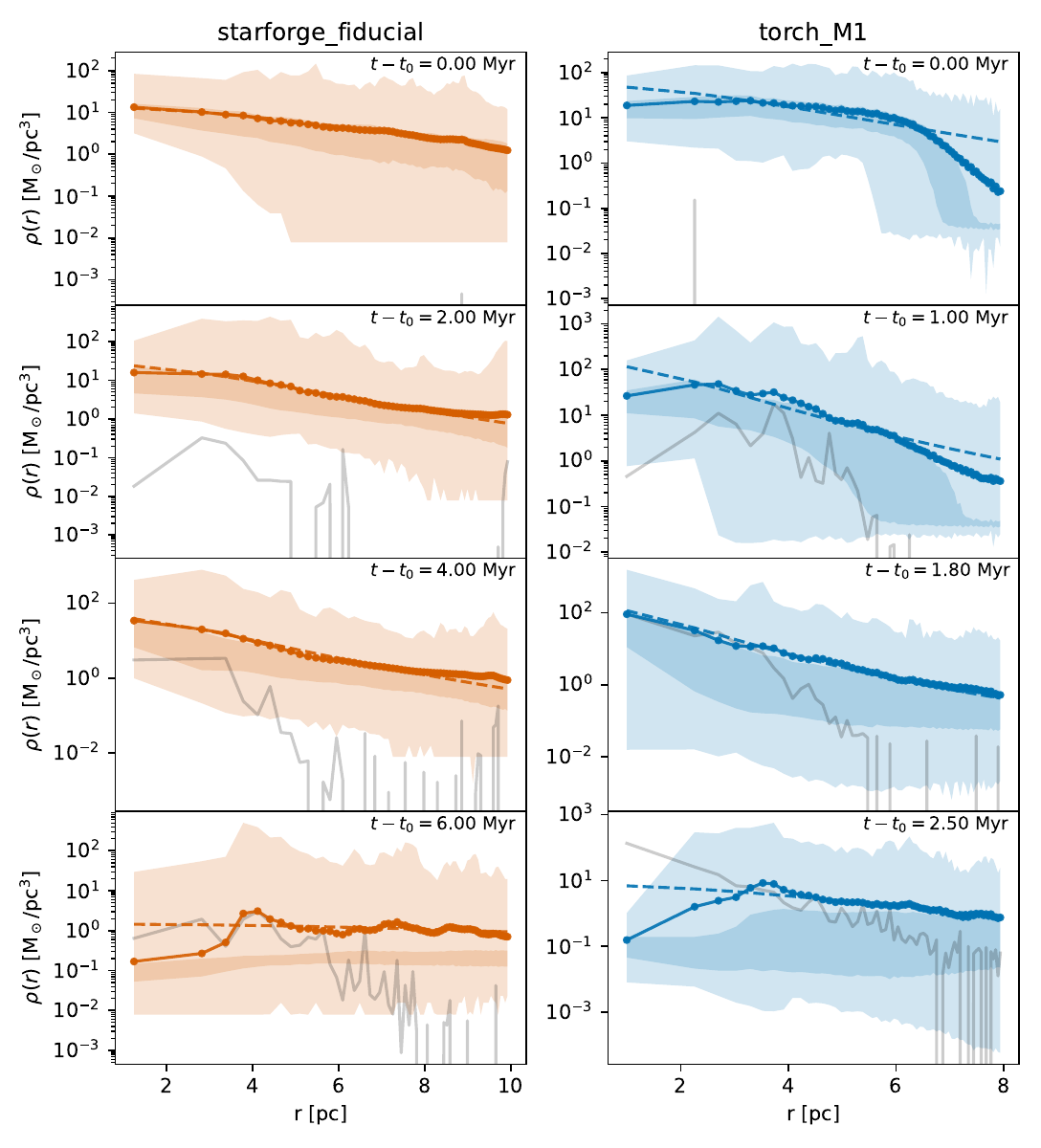}
    \includegraphics[width=0.49\textwidth]{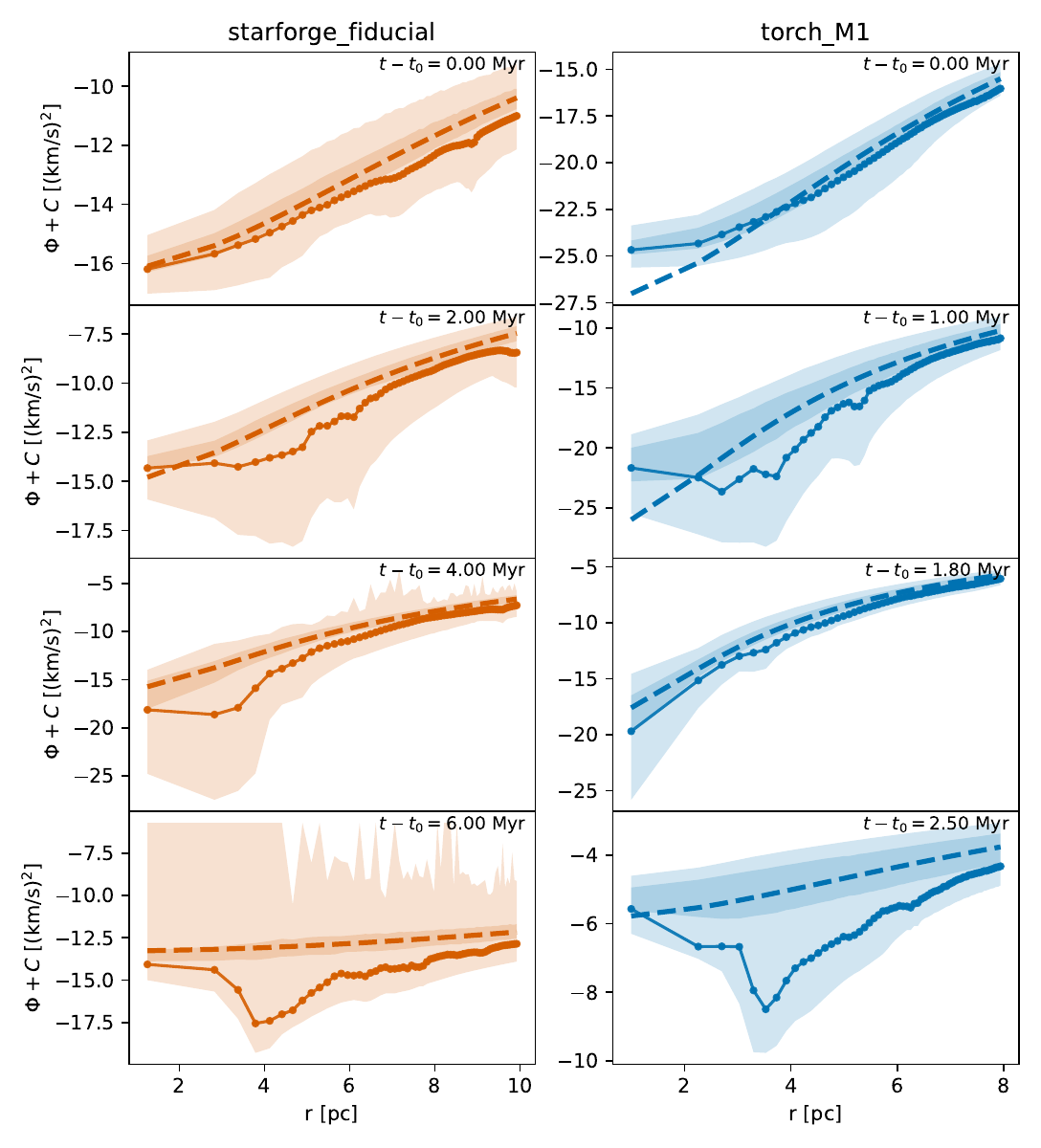}
    \caption{\
    Left: Radial gas and stellar density profiles at different times after
    the formation of the first star. Circles show the gas density profiles measured
    in the MHD simulations. Shaded regions indicate the range of values found at
    each radial shell, with light shading representing the minimum and maximum
    values and darker shading the 25th and 75th percentiles. Dashed lines show the
    Plummer models used to characterize the gas distributions. Grey lines show the
    stellar density profiles for reference.
    Right: Gravitational potential generated by the gas distribution,
    measured directly from the MHD simulation outputs. Circles represent the
    shell-averaged potentials, with shaded regions indicating the minimum,
    maximum, and interquartile ranges at each radius. Dashed lines show the
    gravitational potential generated by the fitted Plummer models from the left
    panel. No fit was performed to the potential itself, apart from an additive
    constant, since the absolute value of the potential depends on the boundary
    conditions of the simulations. The potentials generated from the fitted density
    profiles reproduce the median evolution of the simulated gas potential
    reasonably well.
    }
    \label{fig:gas_profiles}
\end{figure*}

Figure~\ref{fig:gas_profiles} shows example density profiles for the
\starforgefiducial\ and \torchclaude\ simulations at different times. At early
times, the profile can deviate from a Plummer form in the outer parts of the
cloud. However, these regions contain little stellar mass and have little
effect on the stellar dynamics. As the simulations evolve, and before the onset
of strong feedback, the Plummer fits generally improve.

At later times, stellar feedback pushes the gas into an expanding shell-like
configuration that a simple analytic profile cannot reproduce accurately. We
still use a Plummer description in this regime because the gas rapidly
approaches a low-density background and its contribution to the gravitational
potential becomes small. Although the density fits still show deviations from
the simulated profiles, the corresponding gravitational potential is reproduced
much better. We verified this directly using the potential stored in the
simulations. This happens because the potential is a spatial integral of the
density field and is therefore less sensitive to small-scale density
variations. This is why a spherically symmetric description remains useful even
when the density field itself is not well described in detail.

We therefore use the Plummer model to describe the gas structure, given by:
\begin{equation}
\rho(r,t)=\frac{3\mpl(t)}{4\pi \rpl(t)^3}
\left(1+\frac{r^2}{\rpl(t)^2}\right)^{-5/2},
\end{equation}
with the Plummer scale radius, $\rpl$, and the Plummer mass, $\mpl$, describing
the gas structure at all times. To connect the gas evolution across snapshots,
we do not fit each density profile independently. Instead, we fit the evolution
of the retained gas fraction, $\fret(t)$, which more directly captures the
redistribution of gas in the region that matters dynamically for the stars. For
a given gas mass at each snapshot, matching $\fret$ leaves $\rpl(t)$ as the main
parameter controlling the central concentration of the gas. We determine
$\rpl(t)$ by minimizing the difference between the measured and modelled
$\fret$, and then determine the corresponding mass normalization across all
snapshots. Appendix~\ref{app:gas_fitting} describes this fitting procedure in
detail. 

\begin{figure}
\centering
\includegraphics[width=\columnwidth]{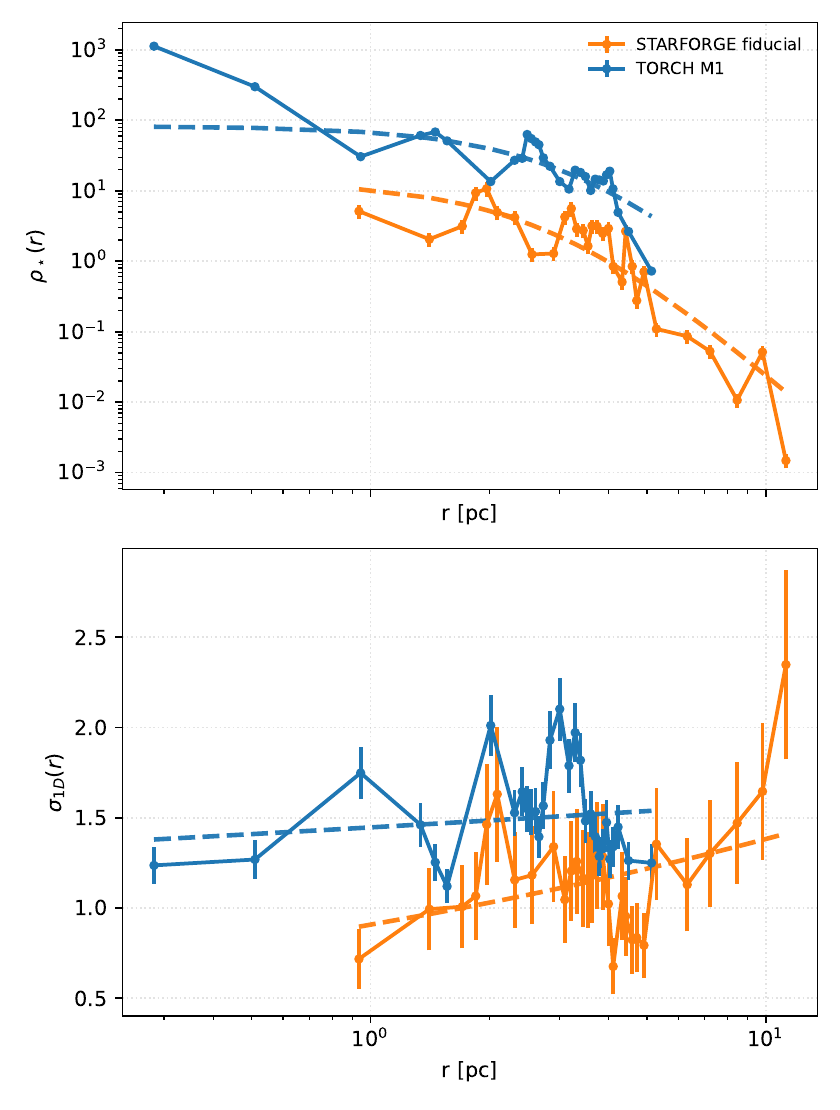}
\caption{
Radial stellar phase-space distributions at the moment stars first appear in
the benchmark MHD simulations. Top panel show the stellar density profiles,
while the bottom panel show the one-dimensional velocity dispersion profiles. The
solid lines show Plummer profile fits to the stellar density and power-law fits
to the velocity dispersion profiles used to define the stellar birth
prescription adopted in this work.
}
\label{fig:starprofiles}
\end{figure}

\subsection{Extracting the phase-space properties of stars at formation}
\label{sec:stellar_extraction}

The properties of newly formed stars reflect both the local and global
conditions of the gas from which they form. However, connecting the gas
structure to the stellar phase-space distribution at birth is not simple, since
it depends on small-scale substructure, time-dependent accretion, and local
feedback. In this work, we therefore use a simpler empirical description of the
global stellar phase-space distribution at birth, and leave a more detailed
treatment for future work. To do this, we record the positions and velocities of
new stars as they first appear in the simulation outputs and use these to build
the birth phase-space distributions. This means that the extracted birth
properties are limited by the time resolution of the simulations.

After collecting these stellar positions and velocities at birth, we construct
the radial profiles of each distribution. Figure~\ref{fig:starprofiles} shows
the results for the \starforgefiducial\ and \torchclaude\ simulations.

We describe the stellar density profile with a Plummer distribution 
\begin{equation}
\rho_\ast(r) = \frac{3 M_\ast}{4\pi R_\text{pl,*}^3}
\left(1 + \frac{r^2}{R_\text{pl,*}^2}\right)^{-5/2},
\end{equation}
where $M_\ast$ is the total stellar mass and $R_\text{pl,*}$ is the stellar
scale radius.

We also measure the one-dimensional stellar velocity dispersion as a function of
radius. In this work, we use a simple centrally peaked profile and approximate
it with a power law,
\begin{equation}
\sigma_{1\mathrm{D}}(r) = \sigma_0
\left(\frac{r}{R_{\rm pl,*}}\right)^{(2-\kappa)/2},
\end{equation}
where $\sigma_0$ is the velocity dispersion at the reference radius and
$\kappa$ controls the radial slope. We use the same scale radius
$R_\text{pl,*}$ inferred from the stellar density profile to keep the phase-space
description consistent. Although the stellar distributions are not truly
isotropic nor perfectly described by Plummer spheres, both the density and
velocity profiles provide remarkably good first-order fits to the newly formed
stellar populations in the benchmark MHD simulations.

In this work, we use these density and velocity dispersion profiles as the
stellar birth prescription. They do not attempt to reproduce the full coupling
between gas substructure and stellar birth, but rather the first-order
phase-space scaling of newly formed stars. The stellar size and velocity scales
are normalized relative to the initial gas cloud at the onset of star formation,
defining the parameters
\begin{equation}
\eta_r = \frac{R_{\rm pl,*}}{R_{\rm pl,gas}(t_0)},
\end{equation}
and
\begin{equation}
\eta_\sigma =
\frac{\sigma_0}{\sqrt{G M_{\rm pl,gas}(t_0)/R_{\rm pl,gas}(t_0)}}.
\end{equation}

For the benchmark simulations analyzed here, the \starforgefiducial\ model gives
$\eta_r = 0.44$ and $\eta_\sigma = 0.28$, while the \torchclaude\ model gives
$\eta_r = 0.55$ and an almost identical velocity scale. We therefore adopt
$\eta_r=0.5$ and $\eta_\sigma=0.3$ as fiducial values for the model grid,
consistent with the approximate scaling found in both benchmark MHD
simulations. Since these ratios are inherited from the star formation process,
they may depend on the adopted star formation prescription and must therefore be
captured at least at this first-order level.

\subsection{Star formation prescription}\label{sec:sf_prescription}

The star formation module, defined by the user, specifies where and how stars
form during the simulation. This prescription may or may not be directly linked
to the evolution of the background gas. For instance, one may couple the star
formation rate to the gas evolution by enforcing a fixed star formation
efficiency per free-fall time, $\epsilon_{\rm ff}$, as commonly assumed in
analytic and semi-analytic models \citep{farias2019,farias2023a}. However, MHD
simulations show that this is not generally the case. Both the global and local
star formation efficiencies evolve in a complex manner, and even in isolated
simulations we observe continuous gas infall prior to the onset of feedback,
which can temporarily increase the local star formation efficiency.

In this work, we adopt a deliberately simple star formation prescription. We
form stars with a fixed phase-space distribution sampled from the empirical
relations described in \S\ref{sec:stellar_extraction}. This choice avoids
introducing additional effects that we aim to isolate in future studies, such as
phase-space substructure inherited directly from the gas, which would require a
more detailed and consistent mapping from the simulations and is beyond the
scope of this first work. We assign stellar masses sampling a \cite{Kroupa2002}
initial mass function (IMF) between 0.08 and 150\,\msun, and we do not include
primordial binaries. This prescription is intentionally minimal and is designed
to ensure that any differences observed in the cluster evolution arise primarily
from variations in the background gas potential rather than from the star
formation model itself.

\subsection{Background gas evolution model}
\label{sec:backgroundgas}

We model the background gas with a time-dependent Plummer sphere. As described
in \S\ref{sec:gas_extraction}, we extract this evolution directly from the MHD
simulations. At each time, we describe the gas with two quantities: the Plummer
mass $\mpl(t)$ and scale radius $\rpl(t)$. We use these in two ways: i) a
\emph{tabulated} evolution, taken directly from the MHD simulations and linearly
interpolated between outputs (see middle panels of Figure~\ref{fig:fret_evol}),
and ii) a parameterized evolution, which we use to explore a wider parameter
space (see right panels of Figure~\ref{fig:fret_evol}). Here we describe the
parameterized version.

The MHD simulations considered in this work all show the same broad behaviour:
an initial contraction phase, during which stars are already forming, followed
by a rapid re-expansion once stellar feedback becomes dominant. We therefore
model the gas evolution with these two phases.

For $t \le t_{\rm ge}$, where $t \le t_{\rm ge}$ is the time when gas expulsion
begins, we describe the collapse phase with
\begin{equation}
\rpl(t) = \rpl{_0}
\left[1-\frac{t-t_0}{\tau_{\rm col}}\right]^{1/2},
\end{equation}
where $\rpl{_0}$ is the scale radius at $t=t_0$. For the gas mass, we use
\begin{equation}
\mpl(t)=M_0+\dot{M}\,(t-t_0),
\end{equation}
where $M_0$ is the gas mass at $t_0$.

For $t>t_{\rm ge}$, we describe the gas expulsion phase with
\begin{equation}
\rpl(t)=\rpl(t_{\rm ge})\,\exp\!\left(\frac{t-t_{\rm ge}}{\tau_{\rm exp}}\right),
\end{equation}
while keeping the gas mass fixed at $\mpl(t_{\rm ge})$. Here $\tau_{\rm exp}$
sets the timescale over which the background potential disappears after the gas has
reached its maximum concentration.

In the runs used in this work, we take $\dot{M}=0$, so that the mass remains
constant and the time dependence enters mainly through $\rpl(t)$. We make this
choice because gas accretion from larger scales during the collapse phase makes
it difficult to define a unique physically meaningful mass slope within the
reference region.

We determine the parameters of this model from the shell-based retention
analysis described in \S\ref{sec:gas_extraction}. We use the onset properties to
fix $\rpl{_0}$ and $M_0$, and we use the peak of the retained gas fraction at the
chosen calibration shell to define $t_{\rm ge}$ and the maximum retention factor
$F_{\rm ret,max}$. In this work, we take the calibration shell to be the
initial 10\% gas Lagrangian radius. For $\dot{M}=0$, the value of
$F_{\rm ret,max}$ fixes the Plummer radius at the transition time,
\begin{equation}
\rpl(t_{\rm ge})=
r_{\rm cal}\sqrt{(f_{\rm cal}F_{\rm ret,max})^{-2/3}-1},
\end{equation}
and the collapse timescale then follows from
\begin{equation}
\tau_{\rm col}=
\frac{t_{\rm ge}-t_0}
{1-\left(\rpl(t_{\rm ge})/\rpl{_0}\right)^2}.
\end{equation}
We then determine the remaining timescale $\texp$ by comparing the
post-peak retention tail of the analytic model to the measured retention curves
and selecting the value that minimizes the residual. The resulting best-fitting
parameters for the benchmark simulations are listed in
Table~\ref{tab:benchmarks}, while the corresponding fits are shown in the right
panels of Figure~\ref{fig:fret_evol}. Despite the simplicity of this
prescription, the parameterized evolution reproduces the global gas evolution of
the MHD simulations remarkably well, capturing both the contraction phase and
the subsequent gas-expulsion phase with only a small number of parameters.

\begin{figure*}
    \centering
    \includegraphics[width=\textwidth]{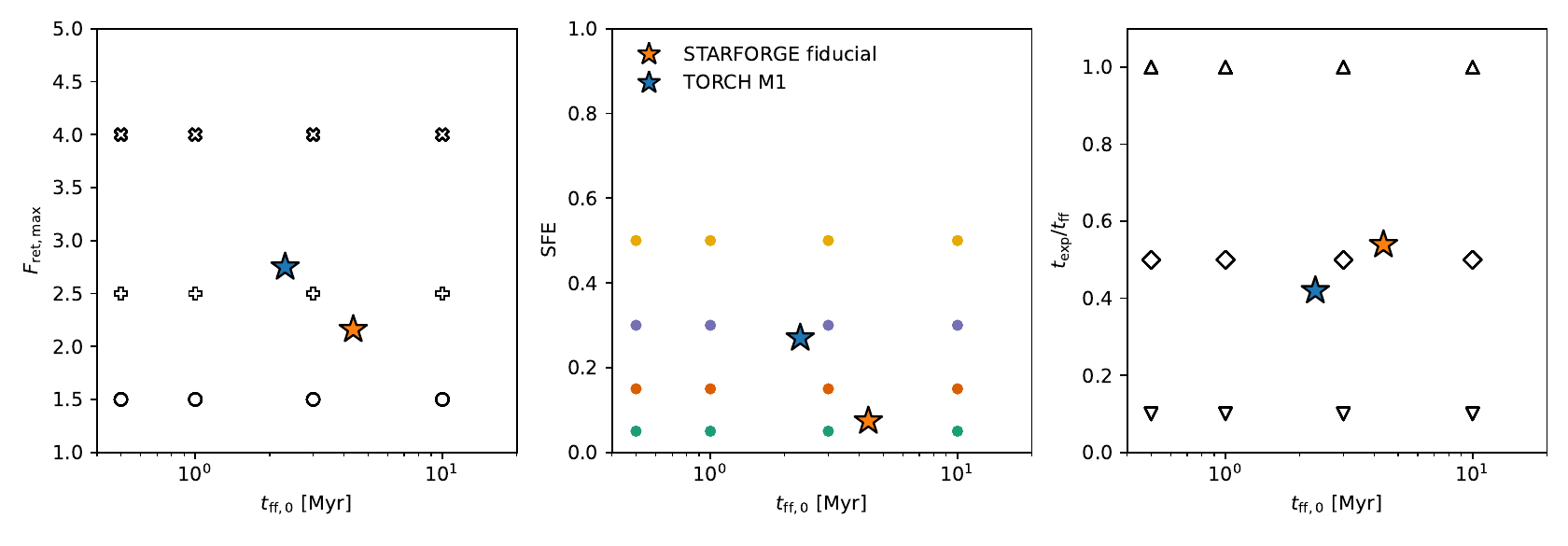}
    \caption{ Parameter-space coverage of the simulation suite. Each panel shows
    the grid (blue circles) explored in this work, with the
    free-fall time $t_{\rm ff}$ on the horizontal axis. Left: Maximum
    retained gas fraction $\fretmax$ versus $\tff$. Middle:
    Star formation efficiency (SFE) versus $\tff$. Right: Gas
    expulsion timescale normalized by the free-fall time, $\texp/\tff$, versus
    $\tff$. Starred symbols indicate the reference MHD simulations (STARFORGE
    and TORCH), placed at their corresponding measured parameter values. The
    grid represents a controlled experimental design intended to show the
    baseline roles of gas concentration, star formation efficiency, and gas
    removal timescale in setting early cluster evolution and stellar kinematics.
    The same color coding and symbols are used through the paper.
    }
    \label{fig:param_space}
\end{figure*}

\subsection{N-body star formation simulations}
\label{sec:sims}

For each target simulation, we run two classes of models. First, we run a
\emph{tabulated} model in which the time-dependent Plummer mass $\mpl(t)$ and
scale radius $\rpl(t)$ are taken directly from the empirical fits to the MHD
simulations and interpolated in time. These runs provide the most direct
comparison to the original simulations, since they add no further assumptions
about the gas evolution.

Second, we run a family of \emph{parameterized} models in which the background
gas follows the analytic description introduced in \S\ref{sec:backgroundgas}.
In these runs, we match the same reference radius $R_{\rm ref}$ and enclosed
gas mass $M_{\rm gas}(<R_{\rm ref},t_0)$ as in the target simulation at the
onset of star formation, while allowing the maximum retained gas fraction
$F_{\rm ret,max}$ to vary. This lets us test how changes in the maximum gas
concentration affect the subsequent stellar dynamics.

For each target simulation, we run multiple realizations with different random
seeds. We choose the number of realizations so that the combined stellar
population in each parameter set is comparable to the stellar mass formed in the
corresponding MHD simulation, while keeping the stellar mass of each individual
realization fixed. We do not include primordial binaries or stellar evolution in
these runs.

We then extend the analysis beyond the benchmark MHD models by exploring a grid
in the main parameters controlling the background gas evolution. Each simulation is
defined by the maximum retained gas fraction $\fretmax$, measured at the
initial 10\% gas Lagrangian radius, the global star formation efficiency
(SFE), the gas-expulsion timescale normalized to the free-fall time
$\texp/\tff$, and the initial free-fall time $\tff$, which
sets the overall density scale of the system.

Across the grid, we explore different values of \fretmax, SFE, $\texp/\tff$, and
$\tff$, extending the ranges covered by the benchmark simulations. We summarize
the parameter mapping in Table~\ref{tab:grid} and show it in
Figure~\ref{fig:param_space}, together with the location of the reference MHD
simulations. With this grid, we aim to isolate how the large-scale evolution of
the background potential affects the stellar dynamics across different
environments, while keeping the stellar birth prescription fixed.

The reference MHD simulations are not necessarily expected to follow the trends
defined by the controlled grid exactly. The grid isolates the effect of the
large-scale evolution of the background potential under spherical symmetry,
while the original MHD simulations additionally contain substructure,
anisotropic collapse, local non-equilibrium motions, and different treatments
of binaries and close encounters. Therefore, deviations between the benchmark
MHD models and the controlled grid are expected and provide information on
which aspects of the stellar dynamics are driven mainly by the global gas
evolution and which depend more strongly on the additional physics present in
the full MHD calculations.

\begin{table*}
\centering
\caption{
    Benchmark D-CAF models used to reproduce the STARFORGE fiducial and TORCH M1
    reference simulations. The tabulated runs use the fitted time-dependent gas
    evolution extracted directly from the MHD simulations, while the parameterized
    runs use the corresponding analytic gas model. In both cases, the stellar birth
    distribution is fixed by the values of $\eta_r$ and $\eta_\sigma$, which set the
    scale of the stellar Plummer distribution relative to the initial cloud.
}
\label{tab:benchmarks}
\begin{tabular}{l l c c c c c c c c c c c}
\hline
Model & Gas & $F_{\rm ret,max}$ & $t_{\rm ge}$ & $t_{\rm exp}$ & $t_{\rm ff}$ & $M_\star$ & SFE & $R_{{\rm pl},\star}$ & $\sigma_0$ & $\eta_{\rm r}$ & $\eta_{\sigma}$& $N$ \\
 &  &  & (Myr) & (Myr) & (Myr) & ($M_\odot$) &  & (pc) & (km\,s$^{-1}$) &  \\
\hline
SF\_fiducial tabulated     & table    & 2.17 & --   & --   & 4.45 & 1023 & 0.074 & 2.94 & 1.11 & 0.44 & 0.28 &10 \\
SF\_fiducial parameterized & analytic & 2.16   & 3.96 & 2.35 & 4.45 & 1023 & 0.074 & 2.94 & 1.11 & 0.44 & 0.28 & 10 \\
TORCH\_M1 tabulated        & table    & 2.92 & --   & --   & 2.36 & 6640 & 0.27  & 3.42 & 1.52 & 0.55 & 0.28 & 2 \\
TORCH\_M1 parameterized    & analytic & 2.75   & 1.60 & 0.97 & 2.36 & 6640 & 0.27  & 3.42 & 1.52 & 0.55 & 0.28 &  2 \\
\hline
\end{tabular}
\tablecomments{
The tabulated \fretmax\ values correspond to the measured peak within the $10\%$ reference radii.
}
\end{table*}

\begin{table}
\centering
\caption{
Parameter ranges adopted for the D-CAF grid. The grid is defined by fixed stellar masses and by varying the gas-evolution and star-formation parameters over the ranges listed below.
\label{tab:grid}}
\begin{tabular}{l l}
\hline
Parameter & Value \\
\hline
$M_\star$ & 1000 $M_\odot$ \\
SFE & [0.05, 0.15, 0.3, 0.5] \\
$F_{\rm ret,max}$ & [1.5, 2.5, 4.0] \\
$t_{\rm ge}/t_{\rm ff}$ & 1.0 \\
$t_{\rm exp}/t_{\rm ff}$ & [0.1, 0.5, 1.0] \\
$t_{\rm ff}$ & [0.5, 1.0, 3.0, 10.0] Myr \\
$\eta_\sigma$ & 0.3 \\
$\eta_{\rm radius}$ & 0.5 \\
$\kappa$ & 1.8 \\
$M_{\rm cl}$ & $M_\star/{\rm SFE}$ \\
$R_{\rm cl}$ & $\left(8GM_{\rm cl}t_{\rm ff}^2/\pi^2\right)^{1/3}$ \\
\hline
\end{tabular}
\end{table}

\section{Results}
\subsection{Reproducing the MHD gas environment}
\label{sec:results_validation}

\begin{figure*}
    \centering
    \includegraphics[width=0.49\textwidth]{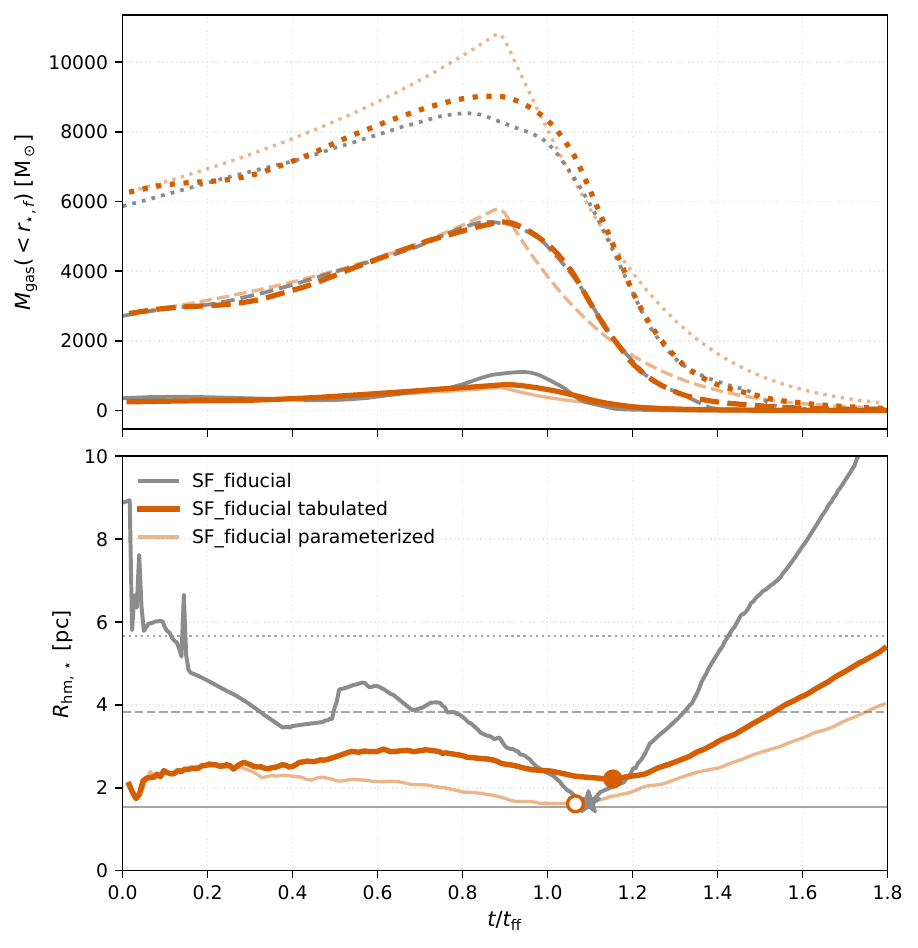}
    \hfill
    \includegraphics[width=0.49\textwidth]{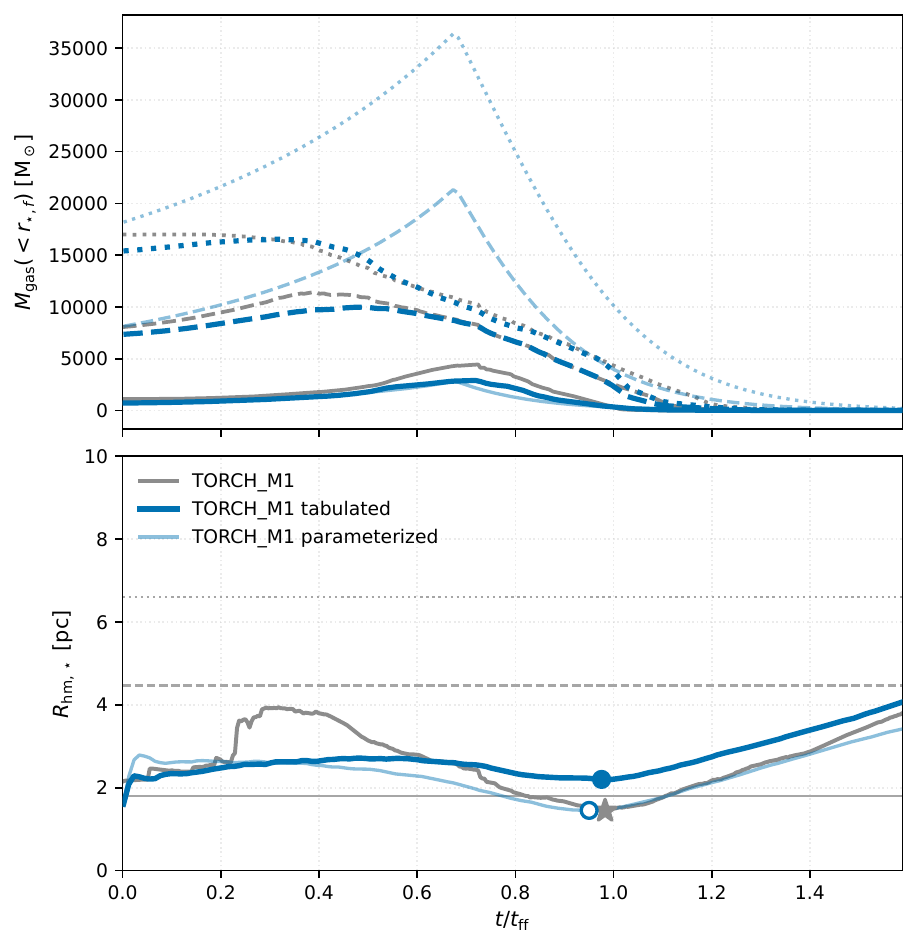}
    \caption{
    Summary of the target simulations designed to reproduce the background
    environments of the STARFORGE fiducial model (left) and the TORCH M1 model
    (right). The bottom panels show the stellar half-mass radius, $R_{\rm
    hm,\star}$, for the original simulation, a realization using the tabulated
    Plummer background gas fits directly (tabulated), and a realization using a
    Plummer sphere with an analytical prescription for its evolution
    (parameterized). The horizontal lines mark the reference radii containing
    10\% (solid), 50\% (dashed), and 75\% (dotted) of the target stellar
    population, obtained by fitting the positions and velocities of stars at
    birth in the original simulations (see text). The top panels show the gas
    mass enclosed within these reference radii. Star symbols highlight the
    location of the minimum stellar half mass radius of the benchmark MHD
    simulations, while filled and empty circles shows the same metric for the
    tabulated and parameterized simulations respectively.
    }
    \label{fig:model_benchmark}
\end{figure*}

As discussed in Section~\ref{sec:gas_extraction}, the Plummer-based
reconstruction of the gas distribution reproduces the large-scale gravitational
potential of the MHD simulations reasonably well. Since the stellar dynamics are
mainly sensitive to the global gravitational field, rather than to the detailed
gas density structure, this is a suitable description of the background
environment in the present models.

We now examine the resulting dynamical evolution of the stellar systems. As a
first step, we compare the target models to the benchmark MHD simulations used
to calibrate the gas potential. Figure~\ref{fig:model_benchmark} shows the
evolution of the stellar half-mass radius in the original MHD simulations
together with the two versions of the target models: the runs using tabulated
background gas and the runs using an analytical parameterized gas evolution. In
both cases, the stars are formed from the same statistical phase-space
distribution, obtained by fitting a Plummer profile to the stellar positions and
velocities at birth in the MHD simulations, as described in
Section~\ref{sec:stellar_extraction}. The top panels show the gas mass enclosed
within fixed reference radii defined from the target stellar population, and are
therefore the same within each set.

We find that both model realizations reproduce the main contraction of the gas
and the corresponding response of the stars. In both the STARFORGE and TORCH
benchmarks, the models reach a similar minimum stellar half-mass radius at
approximately the correct time, after which the systems expand as gas
removal proceeds. This shows that the simplified background descriptions already
capture the main global response of the stellar system.

At the same time, the two implementations do not reproduce the collapse to the
same degree. In both benchmark suites, the minimum half-mass radius of the
original MHD simulations is more closely reproduced by the parameterized runs,
while the tabulated models remain systematically less concentrated, despite
following the gas evolution more directly. This difference arises
because the parameterized models retain a deeper large-scale background
potential during the embedded collapse phase. Although the tabulated models can
reach comparable or even larger central gas concentrations, the parameterized
models retain more gas mass at larger radii during the same period. As a
result, the stellar systems evolve inside a globally deeper potential well
during the collapse phase, leading to a stronger contraction.

This agreement should, however, be treated with care. The minimum value of
$R_{\rm hm,*}$ in the MHD simulations may also reflect the intrinsic phase-space
substructure of the newly formed stars, which is deliberately not included in
the present idealized models. For this reason, the closer match between the
parameterized runs and the MHD benchmarks may be partly coincidental. In that
sense, the difference between the tabulated and parameterized values of
$R_{\rm min,*}$ is more useful as an estimate of how sensitive the stellar
collapse is to small variations in the background gas description. The main
point is therefore not that one realization reproduces the benchmark minimum
exactly, but that the large-scale gas evolution already places a strong
constraint on the stellar concentration, and that modest differences in the gas
model can still produce noticeable differences in the depth of the collapse.

The differences between the benchmark simulations and the target models are already
informative. In the \starforgefiducial\ case, the original simulation shows a
faster post-gas-expulsion expansion than either of the corresponding models. By
contrast, the expansion of the \torchclaude\ case is more closely reproduced by
the target simulations. This difference is already visible during the collapse
phase. The \starforgefiducial\ simulation shows a larger half-mass radius at
early times, significantly larger than in the simplified models, despite having
the same statistical phase-space distribution at birth. This is consistent with
earlier star formation occurring farther from the centre of the potential well,
which leads to a stronger subsequent collapse and, once the gas is removed, to
a faster expansion. By contrast, the \torchclaude\ simulation appears to
follow a stellar assembly history closer to the description adopted in the
present models, which may explain the better agreement during the expansion
phase. This suggests that the radial profile alone does not provide a complete
description of the star formation process. Phase-space correlations,
substructure, and assembly history can also be important ingredients of the
stellar birth prescription.

Even so, the comparison shows that the present models capture the main response
of the benchmark simulations. This gives us a baseline for how an idealized
star formation prescription reacts to a more realistic background gas evolution
than is usually included in $N$-body models. The point of this baseline is not
to reproduce all aspects of the original MHD simulations, but to isolate the
effect of the large-scale gas environment on the early structural evolution of
the stellar system. With this reference in place, we can now examine how the
same baseline responds across the broader parameter space explored by the model
grid.

\subsection{Simulations grid}
\begin{figure*}
    \centering
    \includegraphics[width=\textwidth]{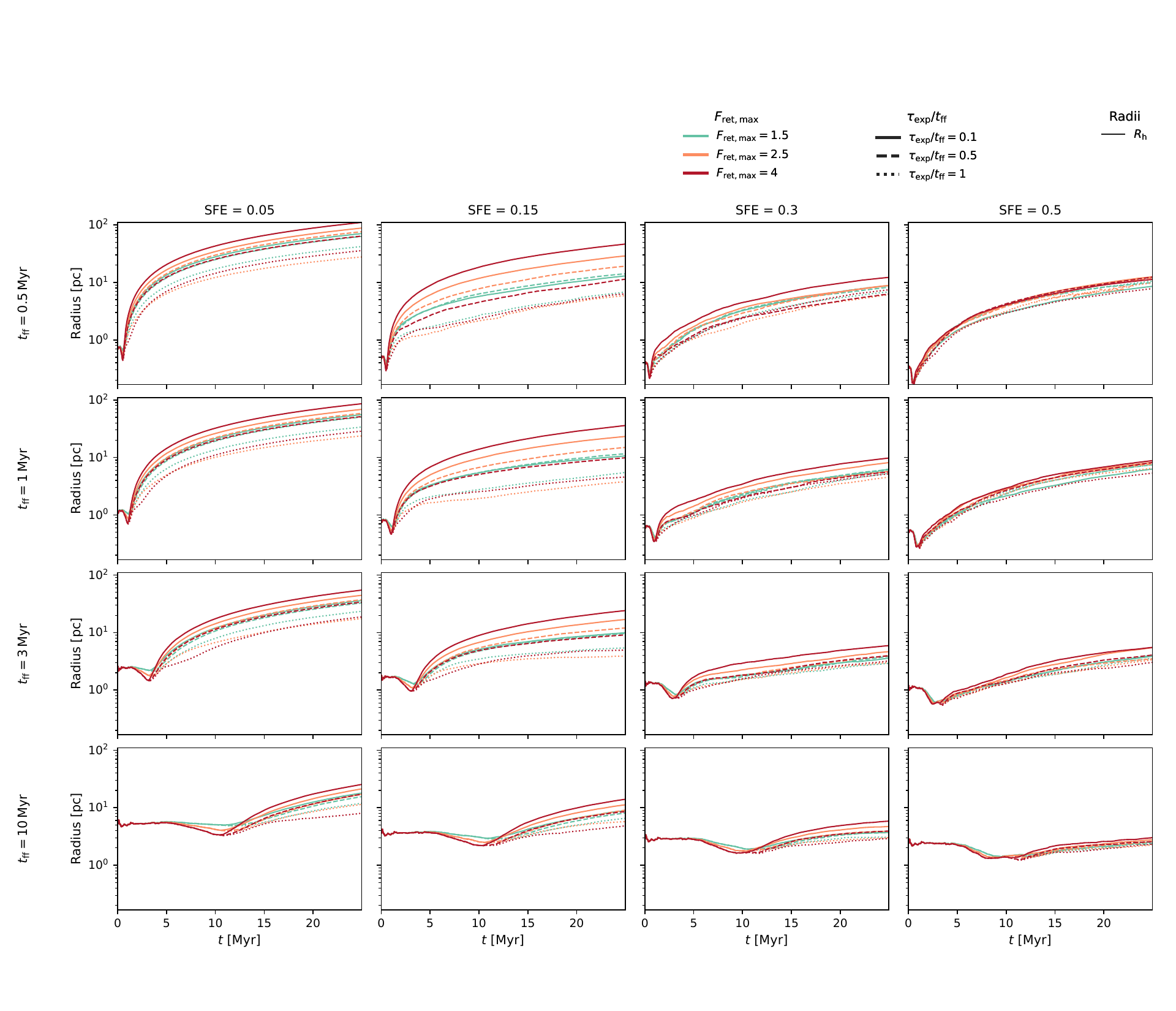}
    \caption{
    Evolution of the stellar half-mass radius, \rhalf, across the simulation grid.
    Columns show different free-fall times, while rows correspond to different
    SFE values. Colours indicate the maximum retained gas fraction, \fretmax,
    and linestyles indicate the gas-expulsion timescale. The overall evolution
    follows a common sequence across the grid: an embedded contraction phase
    while the gas collapses and stars form, followed by expansion after gas
    removal weakens the background potential. Lower-SFE systems and shorter
    free-fall times produce the largest dynamical range in size, reaching both
    deeper contraction and stronger subsequent expansion. Increasing \fretmax
    mainly deepens the contraction, while longer gas-expulsion timescales
    preserve more compact stellar systems after gas removal.
    }
    \label{fig:lagrangian_grid}
\end{figure*}

We now extend the baseline established in the previous section to the wider
parameter space described in Table~\ref{tab:grid}. The goal of this simulation
grid is to explore how stars formed under the same idealized star formation
prescription respond to different background environments, while keeping the
total stellar mass fixed at $1000\,\msun$.

Figure~\ref{fig:lagrangian_grid} shows the evolution of the stellar half-mass
radius, \rhalf, across the simulation grid, grouped by SFE, which sets the depth
of the gas potential relative to the stellar mass, and by \tff, which sets the
density scale of the natal environment. While the trends in these models will
be quantified in the following sections, the figure already provides a global
view of the behaviour of the simulations across the grid. The overall evolution
mirrors the benchmark simulations: an embedded phase during which the stars
form and contract, followed by an expansion phase as gas removal weakens the
background potential. Within this common sequence, \fretmax\ mainly modulates
the depth of the contraction, while the normalized gas-expulsion timescale,
$\tau_{\rm exp}/\tff$, controls how abruptly the systems enter the expansion
phase.

A clear scale contrast is visible across the grid and appears to be mainly set
by \tff. Shorter \tff\ leads to systems with a much larger dynamical range in
size within the first 25\,Myr, reaching both deeper contraction and stronger
subsequent re-expansion. This behaviour is particularly strong in the low-SFE
models. By contrast, larger \tff\ leads to a weaker structural response and
more stable stellar systems. At fixed \tff, increasing SFE generally helps
preserve more compact systems after gas expulsion, while longer
gas-expulsion timescales strongly suppress the later expansion.

Overall, the grid already shows that the global structural evolution is mainly
modulated by \tff\ and SFE, while \fretmax\ and the gas-expulsion timescale
shape the strength of the contraction and expansion phases. However, the
stellar population does not respond to gas removal on the initial cloud
free-fall timescale, but on its own dynamical timescale reached at the onset
of gas expulsion. As the gas contracts during the embedded phase, the stellar
velocity dispersion increases and the stellar crossing time decreases.
Therefore, the same physical gas-expulsion timescale can correspond to very
different effective dynamical regimes depending on the embedded evolution
reached by the system. We now quantify these dependencies directly through the
expansion rates.

\begin{figure*}
    \centering
    \includegraphics[width=\textwidth]{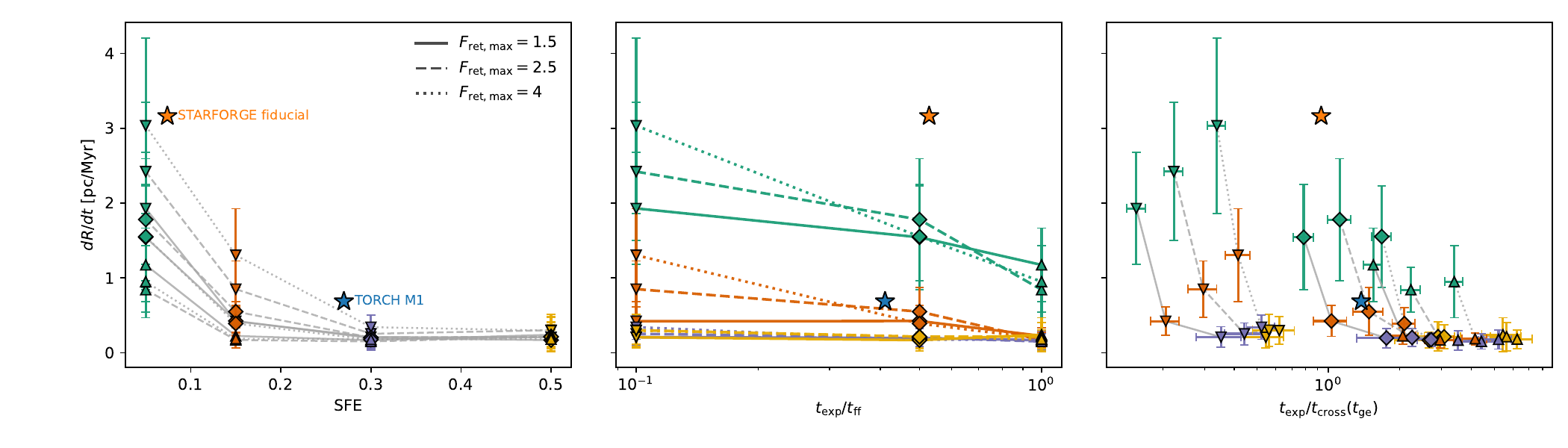}
    \includegraphics[width=\textwidth]{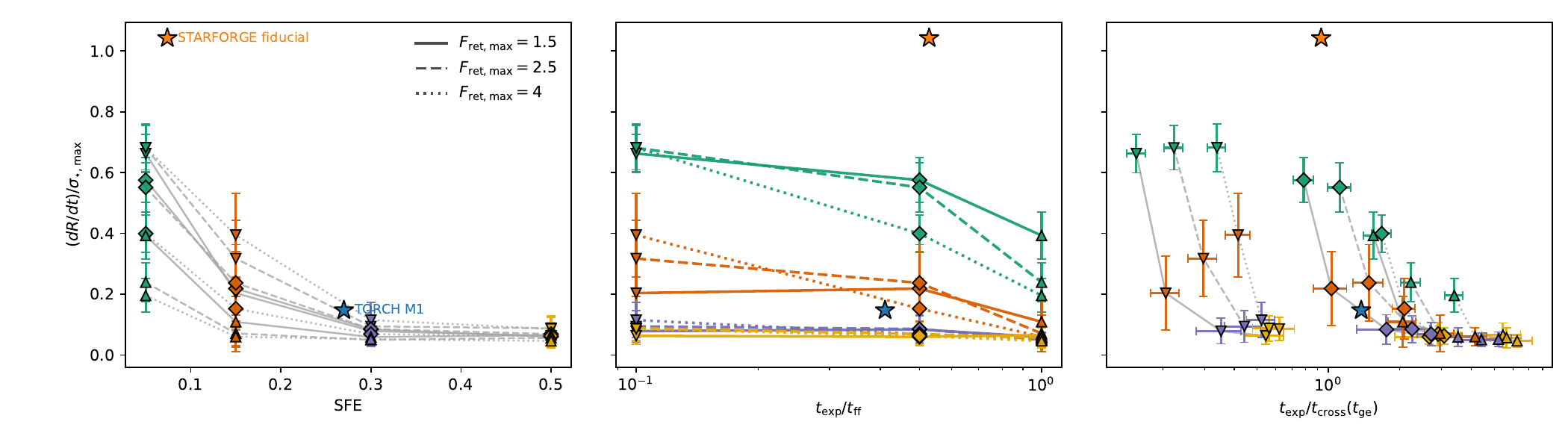}
    \caption{
    Expansion rates measured at 25\,Myr across the full parameter grid. The top
    row shows the expansion rate of the half-mass radius, $\drdt$, while the
    bottom row shows the expansion efficiency relative to the maximum embedded
    velocity dispersion, $\drdt/\sigmamax$. The left column shows the dependence
    on the nominal SFE, the middle column the imposed gas-expulsion timescale
    $\texp/\tff$, and the right column the effective gas-expulsion timescale
    measured relative to the stellar crossing time at the onset of gas
    expulsion, $\texp/\tcross$. Colors and symbols follow
    Fig.~\ref{fig:param_space}, with colors coding the SFE, symbols the gas
    expulsion timescale, and lines connecting families with the same $\fretmax$.
    Filled stars show the location of the reference MHD models. The maximum
    expansion rate reached by a system is set by the embedded velocity
    dispersion before gas expulsion, while the gas-expulsion timescale regulates
    how efficiently this velocity scale is transformed into bulk expansion. The
    dependence on gas-expulsion timescale becomes significantly weaker when
    $\texp/\tcross>1$.\
    }
    \label{fig:drdt}
\end{figure*}

\subsection{Quantifying expansion rates}
\label{sec:expansionrates}

A direct consequence of the embedded evolution described above is that the
stellar systems tend to expand after gas expulsion. However, the expansion rate
is not determined by the imposed gas-expulsion timescale alone, but also by the
stellar dynamical state reached before gas removal and by the fraction of the
stellar population that remains bound. Systems with more gas contraction develop larger stellar velocity dispersions and shorter crossing times prior to
gas expulsion, while the SFE regulates how strongly the gravitational potential
changes once the gas is removed. Both effects determine how efficiently the
embedded velocity scale is transformed into bulk expansion. This makes the
expansion rate a useful quantity to compare across the full simulation grid.

We characterize the expansion using the evolution of the stellar half-mass
radius, \rhalf. After gas expulsion, some systems may still contract for a short
time while they relax, but they eventually enter an expansion phase at a rate
that depends on the model parameters. We therefore fit a straight line to the
time evolution of \rhalf\ and define the slope of that fit as the characteristic
expansion rate, $\drdt$, of the simulation. For cleanly expanding systems this
relation is close to linear, while in systems retaining a larger bound fraction
the evolution of \rhalf\ is noisier and less regular. Even in these cases,
however, the overall evolution can still be described by a mean expansion rate.
In this sense, $\drdt$ should be understood as a global measure of expansion
rather than as an instantaneous quantity. The top panels of
Figure~\ref{fig:drdt} show the recovered values of $\drdt$ across the model
grid. Lower-SFE systems with rapid gas expulsion show the largest expansion
rates, reaching values of up to $\sim 3.5\,\kms$. 

As the SFE increases, the expansion weakens, becoming close to zero for
${\rm SFE}=0.5$. The models separate into three regimes based on the
gas-expulsion timescale. In the impulsive case, expansion is strongest, and the
systems with the largest $\fretmax$ also reach the largest values of $\drdt$. In
this regime, the dependence on $\fretmax$ is clearly visible. When the
gas-expulsion timescale increases to half a free-fall time, which is closer to
the timescales measured in the benchmark MHD cases, the dependence on $\fretmax$
becomes significantly weaker, while SFE still regulates the overall strength of
the expansion. In practice, in this regime only the lowest-SFE systems show
substantial expansion, while most models with ${\rm SFE}\geq0.15$ remain below
$1.5\,\kms$.

This transition becomes clearer when the gas-expulsion timescale is normalized
by the stellar crossing time at the onset of gas expulsion, $\texp/\tcross$.
Systems reaching stronger embedded contraction develop shorter crossing times
before gas removal, moving the same physical gas-expulsion timescale toward a
more adiabatic stellar response. As a result, part of the scatter between
different contraction levels decreases when expressed in terms of
$\texp/\tcross$, suggesting that the stellar response is mainly regulated by the
local stellar dynamical timescale rather than by the initial cloud free-fall
time alone. Remarkably, $\texp/\tcross\sim1$ appears to mark the transition
between an impulsive regime, where the expansion is regulated mainly by the
SFE, and a regime where the gas-expulsion timescale begins to strongly damp the
expansion efficiency.

Additionally, the benchmark MHD models provide some insight into the
limitations of the simplified spherical framework. For instance, the STARFORGE
models tend to reach somewhat larger expansion rates than expected from their
measured \fretmax. This extra expansion may be related to how the stellar
population assembles in the STARFORGE case. As shown in
Fig.~\ref{fig:model_benchmark}, the stellar half-mass radius starts larger and
contracts strongly, in contrast to the TORCH case, where the half-mass radius is
more stable and the collapse is milder. Therefore, when gas expulsion begins in
the STARFORGE case, part of the stellar population may still be falling through
a potential well that is already weakening. The larger expansion rate may then
reflect not only the velocity dispersion reached during formation, but also the
coherent radial structure of the velocity field. This suggests that the
expansion is affected not only by the amplitude of the embedded kinematics, but
also by the assembly history and phase-space coherence of the stellar
population. 

\begin{figure*}
\includegraphics[width=\textwidth]{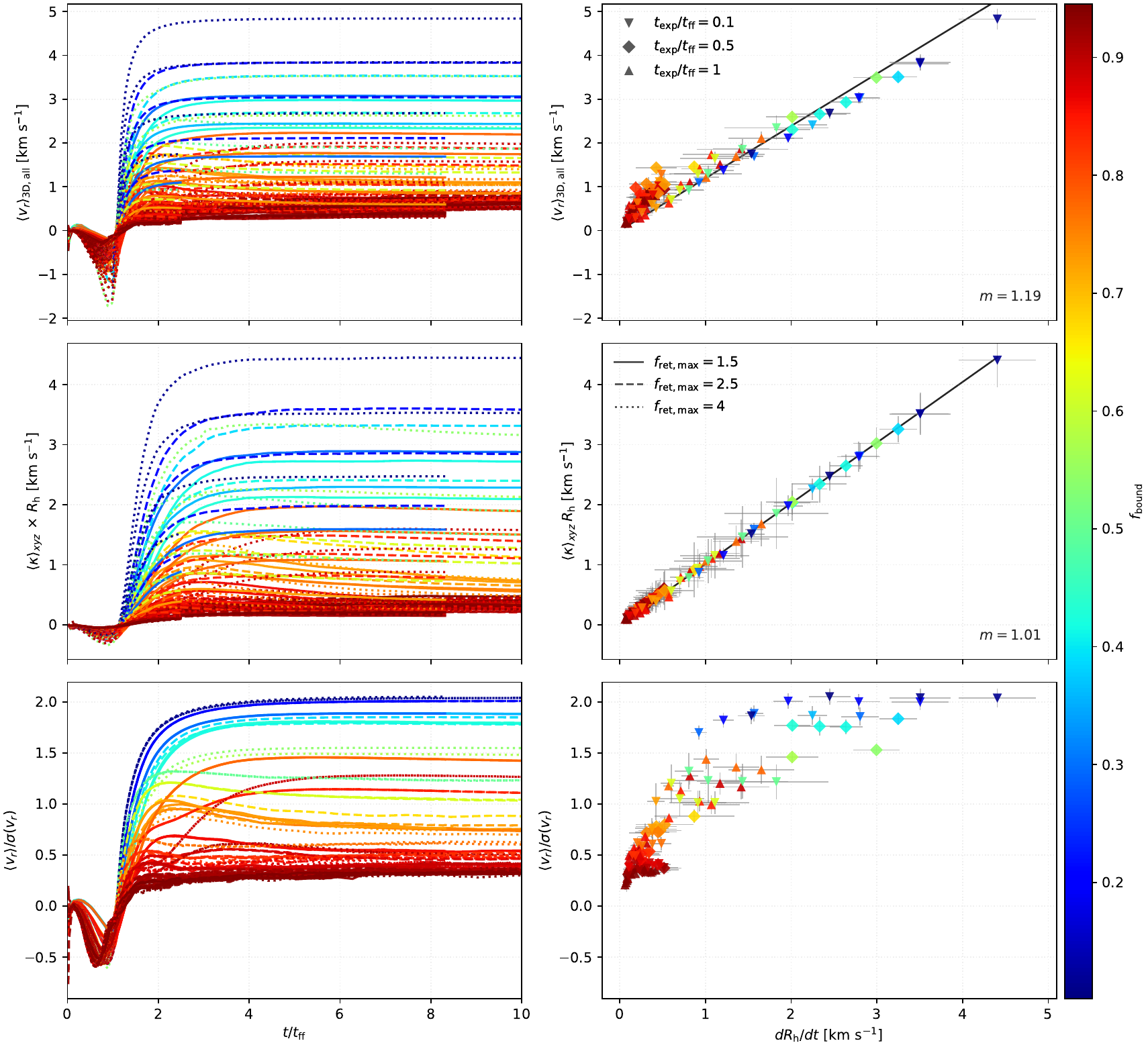}
\caption{
Evolution and interpretation of three observational expansion diagnostics: the
average outward velocity, \vout\ (top panels), the position--velocity expansion
gradient, $\kappa$ multiplied by the current \rhalf\ (middle panels), and the
ratio between outward velocity and radial velocity dispersion, $\vout/\sigmavr$
(bottom panels). The first two quantities directly measure the physical
expansion rate of the stellar population, while $\vout/\sigmavr$ traces how
strongly the velocity field is dominated by coherent outward motion. The right
panels show the evolution of the metrics as a function of free-fall time, with
lines colored by the bound fraction measured $2\,\tff$ after gas expulsion. The
left panels compare the metrics measured at 25\,Myr against the true expansion
rate of \rhalf, with points using the same color scale. Each line and point is
an average of 10 simulations with the same properties, with error bars showing
the standard deviation of the group. Symbols denote the gas-expulsion
timescale following the same coding as Figure~\ref{fig:param_space}, while line
styles code the central concentration \fretmax. After the systems relax,
strongly unbound systems evolve toward more stable values of these metrics,
while systems retaining larger bound fractions show continued evolution and
larger scatter.\
}
\label{fig:expansionmetrics}
\end{figure*}

\subsection{Expansion rate observables} 
\label{sec:observables}

While the gas-expulsion timescale regulates how much expansion is produced, the
SFE is not by itself the quantity that sets the expansion speed. Instead, the
expansion can only draw from the velocity scale already present in the stellar
system when gas expulsion begins. In the models presented here, that scale represents
the maximum stellar velocity dispersion reached before expansion, $\sigmamax$.
The bottom panels of Figure~\ref{fig:drdt} show the expansion rate normalized by
this quantity. In the most impulsive cases, the ratio $\drdt/\sigmamax$ reaches
values of about $0.7$, while it decreases rapidly as either the SFE increases or
the gas-expulsion timescale becomes longer. This shows that $\drdt/\sigmamax$ is
better interpreted as a measure of how efficiently the natal velocity scale is
converted into bulk expansion, rather than as a direct proxy for the natal
velocity scale itself. In this sense, the observed expansion rate provides at
best a lower limit on the velocity scale with which the system emerged from its
embedded phase.

Since the advent of Gaia, several studies have measured signatures of expansion
in young clusters and associations
\citep{kuhn2019,wright2018,quintana2025a,croce2023}. These quantities have
generally been used as indicators of expansion, but their relation to the true
bulk expansion speed of a stellar system is still not well established. Here we
use our homogeneous simulation grid, built from star-formation models, to
compare these commonly used observables directly to the measured expansion of
the systems.

We consider three metrics commonly used in the literature. The first is the
mean outward velocity of the stars, \vout, which has been used as a direct
measure of expansion velocity \citep[e.g.][]{armstrong2026}. The second is the
ratio between outward velocity and radial velocity dispersion,
\vout/\sigmavr\ \citep[e.g.][]{croce2023}. Unlike \vout, this quantity does not
measure the expansion speed directly, but instead traces how strongly the
stellar velocity field is dominated by coherent outward motion relative to
random motions. The third metric is the velocity gradient, $\kappa$, which is
usually measured under the assumption of a single-point expansion pattern and
often interpreted as an expansion timescale.
The difficulty with $\kappa$ is that it is not time-invariant: in an unbound and
isolated expanding system, stellar velocities remain approximately constant
while stellar positions continue to increase, so $\kappa$ changes with time even
when the expansion speed does not. On its own, $\kappa$ is therefore difficult
to compare across systems. Here we instead multiply $\kappa$ by a consistent
size scale, the half-mass radius, so that it is recast from an inverse
timescale into a velocity-like quantity.

Figure~\ref{fig:expansionmetrics} shows the evolution of the three metrics
across the simulation grid. We normalize time by the corresponding \tff. After
gas expulsion, all metrics typically require about two \tff\ to settle into a
roughly constant value. The main exceptions are systems with high bound
fractions ($\gtrsim 50\%$), where continued internal evolution breaks the free
expansion picture expected for stellar associations.

If we compare \vout\ and $\kappa \rhalf$ to the true bulk expansion of the
systems, shown in the right panels of
Figure~\ref{fig:expansionmetrics}, both follow the expansion of the half-mass
radius remarkably well. In the case of \vout, the mean outward velocity
naturally traces the expansion of the stellar distribution. In the case of
$\kappa \rhalf$, the agreement follows from using \rhalf\ itself as the size
scale that converts $\kappa$ into a velocity. We have verified that this
relation is not generally 1:1 for other Lagrangian radii, where an additional
calibration is required.

The behaviour of \vout/\sigmavr\ is qualitatively different. Rather than
tracing the physical expansion rate directly, this quantity measures how clean
the expansion is relative to the internal velocity dispersion of the system.
Strongly unbound systems evolve toward large and approximately constant values
of \vout/\sigmavr, while systems retaining larger bound fractions continue to
evolve and show substantially larger scatter. This suggests that
\vout/\sigmavr\ may provide a useful observational indicator of whether a
stellar population is undergoing clean free expansion or still retains a
significant bound component. Interestingly, both the simulations and the Gaia
cluster sample analyzed by \citet{croce2023} span a similar range of
\vout/\sigmavr\ values at early times, with the distributions appearing to
saturate around values of $\sim2$.

At the same time, applying these relations to observed systems requires a
reliable census of the stellar population, both to estimate \rhalf\ and to
measure \vout\ robustly. In older systems, the outer high-velocity envelope may
already be poorly sampled or no longer identifiable in membership surveys,
biasing both \vout\ and \rhalf\ toward lower values. This effect may become
particularly important once the expanding stellar population becomes spatially
extended and difficult to separate from the Galactic field.

Nevertheless, the relations shown in
Figure~\ref{fig:expansionmetrics} are promising. They still need to be tested
against observational biases and physical effects not included here, such as
anisotropy, primordial binaries, and the Galactic environment. Still, if these
metrics trace the true expansion rate of stellar systems, they open the
possibility of connecting present-day expansion directly to the embedded
velocity scale established during the formation process.

\begin{figure*}
    \includegraphics[width=\textwidth]{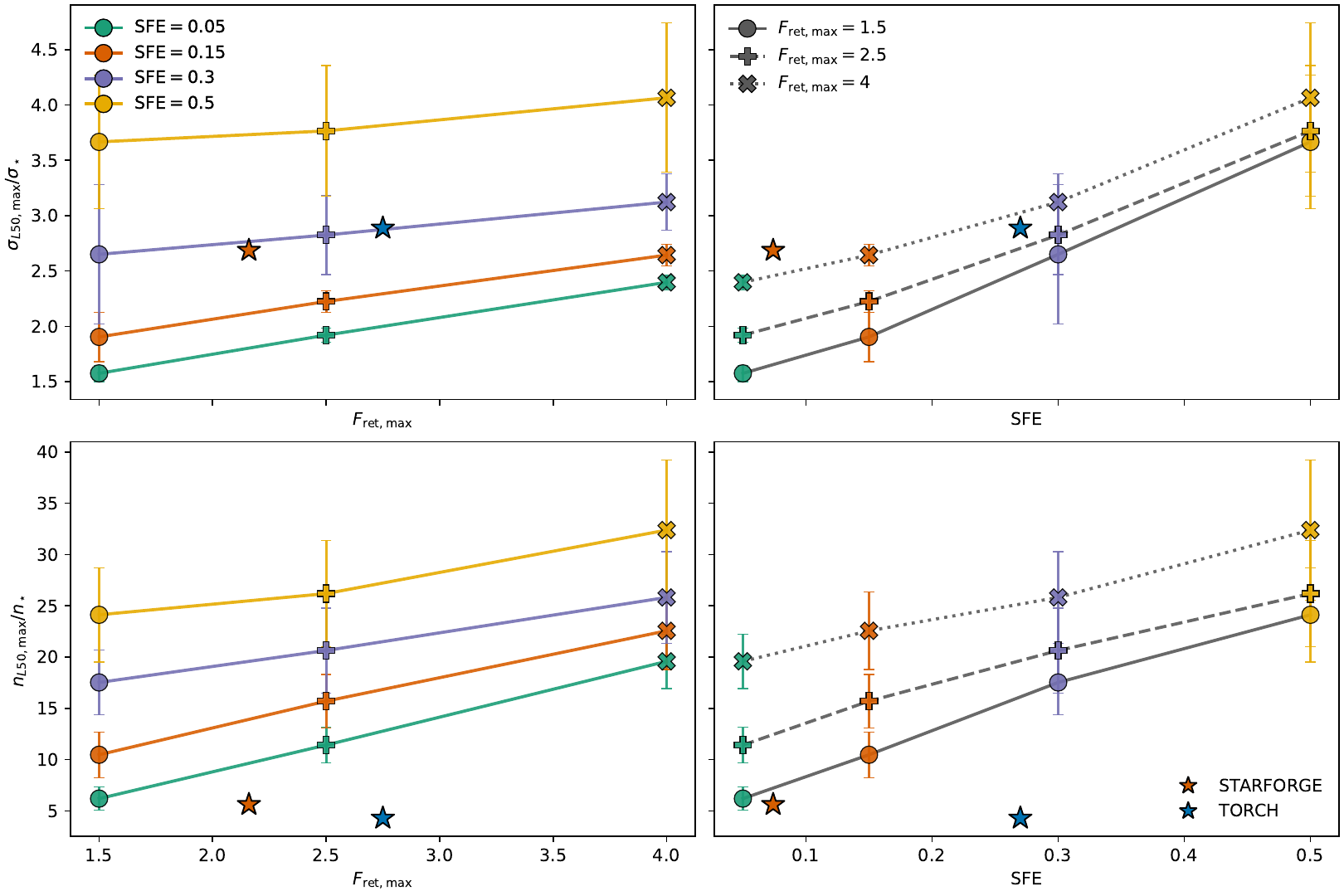}
    \caption{
   Normalized maximum stellar velocity dispersion and maximum half-mass number
   density for stars within \rhalf. The left panels show these quantities as a
   function of $\fretmax$, while the right panels show the same measurements as
   a function of SFE. In all panels, colors indicate SFE and line styles/markers
   indicate $\fretmax$. The normalization factors $\sigma_*$ and $n_*$ are the
   characteristic formation scales defined from the sampled initial distribution.
   }
   \label{fig:focusingfactors}
\end{figure*}

\subsection{ Amplification of stellar kinematics during formation }

As we show in the previous section, the expansion rates are set by the velocity
dispersion reached during the formation phase. It is therefore important to
understand what drives this velocity scale, and how it depends on the details of
the star formation process. Here we examine how this scale is produced across the
grid.
Figure~\ref{fig:focusingfactors} shows the maximum velocity dispersion and
density reached within $\rhalf$, both normalized to their initial values. This
normalization allows us to compare systems with different initial density and
velocity scales on an equal ground, isolating the effect of the formation
process itself.

On the grid of simulations presented here, stars are formed sampling from a
fixed distribution (see \S\ref{sec:sf_prescription}), where the spatial and
velocity scales are set by the parameters $\eta_r$ and $\eta_\sigma$, and tied
to the background gas potential. This allows us to define consistent reference
scales $\sigma_\star$ and $n_\star$ for each model. We take $\sigma_\star$ from
the velocity normalization, $\sigma_\star = \eta_\sigma \sqrt{G M_{\rm pl}/R_{\rm pl}}$,
and define the characteristic stellar density as the mean density within the
half-mass radius,
\begin{equation}
n_\star = \frac{3}{8\pi}\,\frac{N_{\star,\rm final}}{R_{\rm pl}^3},
\end{equation}
where the numerical factors account for the relation $\rhalf \approx 1.3\,\rpl$
and the use of half the stellar population.

With this normalization, we find that both $\fretmax$ and the SFE
systematically increase $\sigmamax$ and $n_{*,\max}$. The dependence on SFE
can be understood from the way the stellar distribution is initialized relative
to the gas potential. At fixed $\eta_\sigma$ and $\eta_r$, the stellar size and
velocity scales are set as fixed fractions of the background gas potential. As a
result, changing the SFE modifies the depth of the potential in which the stars
are embedded without changing their relative scaling to it.

Since the stellar mass is fixed across the grid, lower-SFE systems correspond to
a more massive gas background, and therefore to a deeper potential. The stellar
velocity dispersion then increases accordingly, while the stellar self-gravity
remains unchanged. This produces a systematic variation in the dynamical state of
the stellar component, such that lower-SFE systems are born dynamically hotter,
while higher-SFE systems are closer to equilibrium and can collapse more
efficiently prior to gas expulsion, reaching higher densities and velocity
dispersions.

This behaviour follows directly from the way stars are formed in the underlying
MHD simulations. In that case, the stellar distribution inherits its size and
kinematic scales from the local gas potential at the time of formation, rather
than from the self-gravity of the stellar component alone. The parameters
$\eta_\sigma$ and $\eta_r$ therefore encode this coupling between stars and gas,
and the resulting dependence on SFE reflects the dynamical state in which the
stars are assembled. These trends highlight the subtleties of the dynamical
formation process imprinted on the resulting stellar population, which are often
not captured by MHD simulations alone.

On top of this trend, the evolution of the gas potential modulates the degree of
contraction in a more direct way. Increasing $\fretmax$ leads to a more
centrally concentrated gas distribution, deepening the potential felt by the
stars and driving an additional increase in both the velocity dispersion and the
stellar density before gas removal. 

Figure~\ref{fig:focusingfactors} shows that the amplification of the velocity
dispersion is moderate across the grid, ranging between factors of $\sim 1.5$ to
4 relative to the intrinsic values from the star formation prescription. In
contrast, the corresponding increase in $n_{*,\max}$ is significantly stronger,
reaching factors of $\sim 5$ to 30 above the formation values. At fixed
$\fretmax$, increasing the SFE produces the largest change in both quantities,
while variations in $\fretmax$ at fixed SFE lead to more moderate changes in
$\sigmamax$ and a stronger response in $n_{*,\max}$, particularly at low SFE.
Across the full range of SFE, $\fretmax$ increases the central density by an
average factor of $\sim 2$ between $1.5 < \fretmax < 4$, and a milder average
factor of $\sim 1.3$ in the velocity dispersion. In both cases, low-SFE systems
show the strongest dependence.

The benchmark MHD models do not follow the density trends to the same degree as
the velocity amplification. In particular, the TORCH models reach much lower
density concentrations than expected from their measured \fretmax\ despite
reproducing the velocity amplification more closely. This suggests that the
global depth of the potential is reasonably captured by the simplified
framework, while the spatial concentration of the stellar population is more
sensitive to the detailed assembly process. In particular, the TORCH stellar
distribution remains significantly more structured than a smooth Plummer
profile, such that global density measures based on the half-mass radius may
underestimate the density reached in local collapsing regions. Consistently,
using a local-density estimator substantially increases the measured density
concentration in the TORCH case, although we leave a systematic comparison
between global and local density estimators for future work.

Considering that we explore here only a relatively narrow range of $\fretmax$
around the values found in the MHD simulations analyzed, the central
concentration of gas already plays a significant role in increasing the central
concentration of stars. In addition, the relative scaling of the newborn stellar
population with respect to the gas is an important factor that sets the velocity
scale of the final stellar system, with direct implications for how fast these
systems will expand. It remains to be explored with a broader set of MHD
simulations how these factors vary across different star formation
prescriptions. We note, however, that in other MHD simulations not included
here, $\fretmax$ can reach higher values. Here we have just explored the
consequences of such increase, however the conditions that lead to different
\fretmax\ values and the ranges that are physically possible is a matter that
has not yet been explored.

\begin{figure*}
    \centering
    \includegraphics[width=\textwidth]{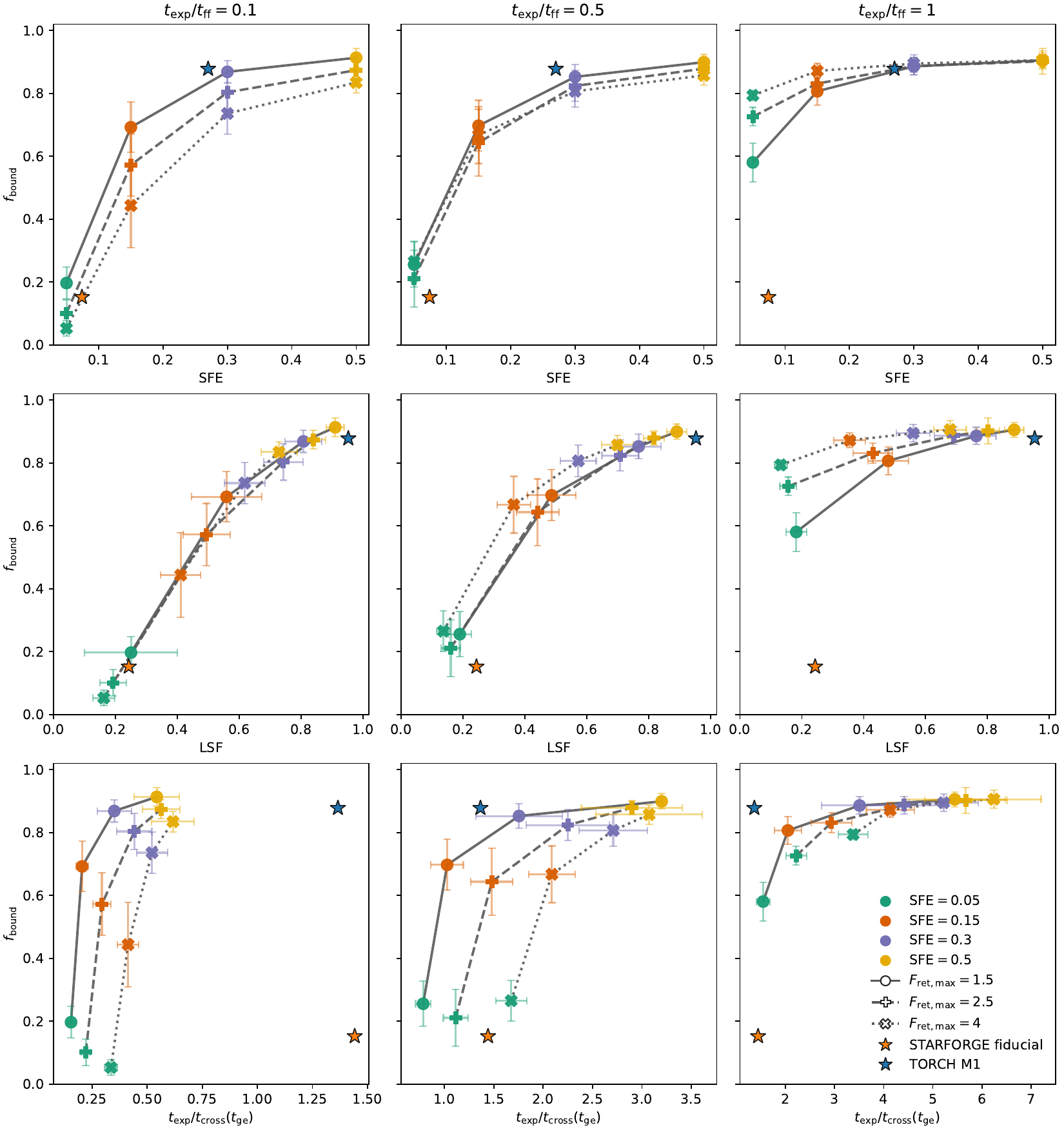} \\
    \caption{\
    Bound fraction measured at $t_{\rm ge}+2\,t_{\rm ff}$ across the simulation
    grid. Columns show the three imposed gas-expulsion timescales,
    $\texp/\tff=0.1$, $0.5$, and $1$, from left to right. The top row shows the
    bound fraction as a function of the nominal global SFE, the middle row as a
    function of the local stellar fraction (LSF) reached at the onset of gas
    expulsion, and the bottom row as a function of the effective gas-expulsion
    timescale normalized by the stellar crossing time at gas expulsion,
    $\texp/\tcross$. Colours represent SFE, symbols connected by lines the
    $\fretmax$, and filled stars show the reference MHD models. The LSF orders
    the impulsive gas-expulsion case, while $\texp/\tcross$ better explains the
    adiabatic regime. 
    }
    \label{fig:boundfraction}
\end{figure*}

\subsection{Bound fraction across the grid}

The gas expulsion phase is a crucial stage in the evolution of young star
clusters, and as MHD simulations increase in complexity it becomes important to
identify the physical quantities that determine whether stellar systems survive
gas removal. Classical studies have shown that the survival of young star
clusters depends primarily on the SFE and the gas-expulsion timescale . A high
SFE leaves enough stellar mass to retain the gravitational potential once the
gas is removed, while slow gas removal gives stars time to adapt to the evolving
potential. On the other hand, rapid gas expulsion leaves many stars unbound
\citep{lada1984,goodwin2006,baumgardt2007}. Other factors are also known to
affect cluster survival, such as substructure, which can locally raise the
stellar density \citep{smith2011,farias2015}, and the dynamical state of the
stars at the time of gas expulsion, with sub-virial systems generally more
likely to survive rapid gas loss \citep{lee2016,farias2018}.

MHD simulations provide important information on how these different processes
compete during the formation phase. In this first work, we isolate the two main
processes explored in classical studies, the SFE and the gas-expulsion
timescale. However, unlike classical idealized models, the gas distribution
contracts while stars are forming and the stellar population is not initially in
equilibrium. As a result, the embedded phase not only changes the stellar
fraction that reaches the center before gas expulsion, but also the stellar
dynamical timescale itself. Systems reaching stronger contraction develop both
larger stellar velocity dispersions and shorter stellar crossing times before
gas removal. Therefore, the response of the stellar population depends not only
on the imposed gas-expulsion timescale, but on how this timescale compares to
the local stellar dynamical time reached during the embedded evolution.

This is where the initial stellar kinematics become important. Systems formed
with colder stellar velocities collapse more efficiently toward the center,
raising the local stellar fraction prior to gas expulsion. In contrast, hotter
systems are less able to follow the contraction of the gas potential, leading to
lower central stellar concentrations and therefore lower bound fractions after
gas removal.  Therefore, the relevant quantity for cluster survival is not
necessarily the global SFE alone, but the local stellar fraction established
during the formation phase. The middle row of
Figure~\ref{fig:boundfraction} shows the bound fraction measured
$2\,t_{\rm ff}$ after gas expulsion as a function of the nominal SFE for two
gas-expulsion timescales.

We recover the classical behaviour where larger SFE and slower gas removal lead
to larger bound fractions. In the impulsive gas-expulsion regime,
$t_{\rm exp}/t_{\rm ff}=0.1$, the bound fraction increases steeply with SFE,
with low-SFE systems becoming largely unbound and higher-SFE systems retaining a
significant bound component. However, at fixed nominal SFE, the models follow
different trends as a function of $\fretmax$. Systems with larger $\fretmax$
follow lower bound fractions trends across the SFE.

As shown in the previous section, increasing $\fretmax$ raises both the stellar
density and the velocity scale reached during formation. In the impulsive
gas-expulsion regime, the increase in velocity dispersion dominates the response
of the system after gas removal, producing dynamically hotter stellar
populations that are more easily disrupted. At the same time, the contraction of
the stellar distribution also modifies the stellar fraction that reaches the
center prior to gas expulsion, making the LSF the parameter that determines
survival. 

The behaviour changes in the slower gas-expulsion regime,
$t_{\rm exp}/t_{\rm ff}=1$, where systems with larger $\fretmax$ now reach
larger bound fractions despite their lower LSF. As discussed in the previous
section, increasing $\fretmax$ raises the stellar velocity dispersion reached
during formation and shortens the stellar crossing time at the onset of gas
expulsion. The same physical gas-removal timescale therefore becomes
effectively more adiabatic from the perspective of the stars, allowing the
stellar population more time to adapt to the evolving potential. As a result,
systems with stronger embedded contraction can retain a larger bound fraction
despite their lower LSF.

The bottom row of Figure~\ref{fig:boundfraction} shows the bound fraction as a
function of $\texp/\tcross$ at the onset of gas expulsion. In the impulsive
regime, $\texp/\tff=0.1$, the different $\fretmax$ families remain clearly
separated, with systems reaching stronger contraction producing systematically
lower bound fractions. In this case, the gas-removal timescale remains shorter
than $\sim0.7\,\tcross$ for all models, such that survival is regulated mainly
by the LSF. In the intermediate regime, $\texp/\tff=0.5$, the separation between
$\fretmax$ families becomes much weaker despite the broad range of stellar
crossing times reached by systems with the same SFE, showing that $\fretmax$ no
longer regulates the bound fraction. In this regime, the SFE becomes the
dominant parameter controlling survival. Finally, in the slowest gas-expulsion
regime, $\texp/\tff=1$, the different $\fretmax$ families collapse approximately
into a single trend as a function of $\texp/\tcross$. This shows that once the
gas-expulsion timescale becomes considerably longer than the stellar dynamical
timescale, the stellar response becomes regulated primarily by the local
crossing time reached during the embedded phase rather than by the initial cloud
free-fall time alone.

Interestingly, the benchmark MHD models appear to lie close to the transition
between the impulsive and adiabatic regimes. Their nominal \texp/\tff\ is close
to $0.5$, while their measured \tcross\ at the onset of gas expulsion is also
close to unity, placing them in the regime where the global SFE becomes the
main parameter regulating survival. In this context, the low-SFE STARFORGE
models naturally reach low bound fractions, while the TORCH models lie in the
high-bound-fraction regime where the dependence on both \fretmax\ and
$\texp/\tcross$ becomes significantly weaker. However, this balance may shift on
different cloud mass scales depending on how efficiently stellar feedback is
able to remove the gas from the system.

\section{Discussion}

Star cluster formation simulations appear to show a common global behaviour. As
star formation begins, gas continues collapsing toward the centre of the cloud.
This collapse transfers momentum to the newly forming stars while at the same
time increasing the central concentration of gas, such that during the embedded
phase stars are assembled inside an evolving background potential. This process
continues until stellar feedback becomes strong enough to reverse the gas flow
and drive the cloud expansion. We have shown that this behaviour is consistent
in the two MHD models used here as reference, but we have also explored other
MHD simulations showing similar trends, whose analysis will be presented in a
future work.

In agreement with seminal studies such as \cite{baumgardt2007} and many works
that followed, we find that the global structure of the gas during the embedded
phase can still be reasonably described by a smooth Plummer-like potential.
However, instead of remaining static, the gas distribution contracts while stars
are forming, continuously modifying the stellar dynamical state prior to gas
expulsion. As gas collapses toward the centre, the stellar fraction decreases
while the stellar velocity scale and central concentration increase. The
embedded stellar population is therefore not evolving independently from the
gas, but dynamically coupled to the global evolution of the cloud.

This behaviour has important consequences for both the survival of young star
clusters and the expansion of stellar systems after gas expulsion, all strongly
modulated by the gas-expulsion timescale. In the impulsive gas-expulsion regime,
the later evolution is determined primarily by the local stellar fraction. As
gas collapses, its contribution to the binding potential increases, raising the
stellar velocity dispersion before rapidly vanishing. The stellar population is
then left in a highly supervirial state, where the expansion rate depends both
on how many stars remain bound and on the velocity scale reached during the
embedded phase. In this regime, systems reaching larger LSF survive gas
expulsion more efficiently.

Interestingly, this behaviour changes once gas expulsion becomes slower. Faster
gas contraction decreases the LSF, but at the same time shortens the dynamical
timescale of the stellar population by increasing the stellar velocity
dispersion. As a result, gas removal becomes effectively more adiabatic from the
perspective of the stars in systems reaching stronger contraction. The MHD
simulations analyzed here suggest that gas removal occurs closer to the
adiabatic regime rather than as an instantaneous event, requiring roughly half
to one free-fall time to effectively clear the gas
\citep{farias2023,wainer2025}. While these timescales may depend on cloud scale
and feedback processes \citep{dale2015,dinnbier2020}, classical studies have
already shown that cluster survival depends primarily on the gas-expulsion
timescale measured relative to the stellar crossing time
\citep{lada1984,geyer2001,baumgardt2007}. In particular,
\citet{geyer2001} showed that more concentrated systems survive gas expulsion
more efficiently because of their shorter dynamical timescales. In this work, we
show that the embedded evolution naturally produces these shorter dynamical
times by concentrating gas toward the centre and dynamically heating the stellar
population during formation.

Remarkably, both benchmark MHD models lie close to
$\texp/\tcross\sim1$, near the transition between the impulsive and adiabatic
regimes identified in this work. In this regime, the global SFE becomes the
main parameter regulating survival, while the dependence on the embedded
contraction weakens significantly. It remains to be explored whether other MHD
simulations populate the same regime. In this picture, the feedback strength
relative to the binding energy of the cloud is not the only quantity governing
gas expulsion. The dynamical evolution of the stellar population and its ability
to shorten its own dynamical timescale during the embedded phase become equally
important, and it is still uncertain which regime is more representative of
Galactic star-forming regions.

The heating of the central stellar population also has important consequences
for the later expansion of young stellar systems. We have shown that the
velocity dispersion reached during the embedded phase is later transformed into
bulk expansion motions. However, this transformation is not fully efficient, as
part of the embedded kinetic energy can remain stored in bound motions or be
redistributed during the relaxation that follows gas expulsion. The efficiency
with which this embedded velocity scale is converted into expansion therefore
depends both on the boundness of the stellar system and on the gas-expulsion
timescale. Current interpretations of expansion in stellar associations remain
largely qualitative rather than quantitative. Most expansion diagnostics are
designed to determine whether a stellar system is expanding or estimating
approximate expansion timescales
\citep{wright2018,wright2019,kuhn2019,armstrong2026}, while the connection between
these observables and the physical expansion rate of the stellar population has
received less attention.

We have shown here that two commonly used expansion metrics are directly
connected to the true expansion rate of the stellar population, provided the
full kinematic information is available. This opens the possibility of using the
expansion of largely unbound systems to recover information about their embedded
phase, at least as a lower limit, once these observables are properly calibrated
against observational biases. For instance, \citet{kuhn2019} measured outward
motions in a large sample of young stellar associations, generally finding
modest expansion velocities. Similar results have also been found in other Gaia
studies of nearby stellar systems \citep{armstrong2026,croce2023}. Considering
that both theoretical and observational evidence point toward low global SFE,
such slow outward motions could suggest that gas expulsion was relatively slow
compared to the stellar dynamical timescale, reducing the efficiency with which
the embedded velocity dispersion is converted into bulk expansion. On the other
hand, these low expansion rates may also be affected by observational biases and
incompleteness \citep{buckner2023}, since the fastest and most extended stellar
populations are more difficult to identify and a significant fraction of the
stellar envelope may already be missing from current samples.

A recent Gaia survey of expansion signatures in nearby young stellar clusters by
\citet{croce2023} showed that younger systems tend to display stronger
expansion signatures than older clusters. In our simulations, however, strongly
unbound systems preserve nearly constant expansion metrics over long timescales,
while the only clusters that show significant evolution in these metrics are
the ones that expand more slowly and contain a larger fraction of bound stars.
This suggests that part of the observed weakening of expansion signatures with
age may not necessarily reflect the true expansion history of the
population, but instead the increasing difficulty of recovering the full stellar
envelope at later times, leaving mainly the denser bound component observable.
Under this picture, there may exist a limited age window where expansion
measurements provide the clearest information about the embedded velocity scale
of the stellar population before the expanding envelope becomes too diluted
observationally. Following the results of \citet{croce2023}, this window
appears to close at ages of $\sim30\,$Myr.

This interpretation, however, remains to be tested against observational biases
and a more realistic star formation process that must include primordial
binaries, stellar evolution and, importantly, substructure. All these processes
may contribute to modulating the expansion signatures and bound-cluster
thresholds, and must be explored in future work. In particular, the comparison
with the TORCH models suggests that stellar density amplification can depend
strongly on how the stellar concentration is measured. While the global velocity
amplification is reproduced reasonably well by the simplified framework, global
density estimators based on half-mass quantities may underestimate the density
reached in local collapsing regions when strong substructure is present.
Therefore, the grid of models presented here serves as a baseline to isolate the
fundamental dynamical consequences that modern MHD simulations imprint on the
stellar populations they produce. In this sense, connecting realistic embedded
environments to long-term stellar dynamics becomes necessary to identify which
properties of the star formation process are actually preserved in the
observable kinematics of young stellar systems.

\section{Conclusions}

In this paper, we presented the Dynamical Cluster Assembly Framework (D-CAF), a
framework designed to connect embedded star formation histories from MHD
simulations to the long-term stellar dynamics of young stellar systems. Using
this framework, we constructed a grid of controlled $N$-body models with
evolving background gas potentials to examine how the evolution of the gas and
stellar population prior to gas expulsion regulates both the survival and
expansion of young stellar systems after gas removal.

Our main results can be summarized as follows:

\begin{itemize}

\item The embedded phase is not well represented by a static background
potential. Instead, the gas contracts while stars are forming, continuously
modifying the stellar dynamical state prior to gas expulsion. As gas collapses
toward the centre, the stellar velocity scale and central concentration
increase, while the stellar fraction within the embedded region decreases.

\item The survival of stellar systems after gas expulsion is strongly regulated
by the gas-expulsion timescale. In the impulsive regime, systems become highly
supervirial and their later evolution depends primarily on the local stellar
fraction reached during the embedded phase. In contrast, once gas removal occurs
on timescales comparable to the stellar dynamical time, the response of the
stellar system becomes strongly damped and the expansion weakens substantially.

\item In the adiabatic gas-expulsion regime, gas contraction shortens the
stellar dynamical timescale prior to gas removal. As a consequence, for a fixed
gas-expulsion timescale, systems with stronger contraction experience gas removal in a more
adiabatic regime relative to their stellar crossing time. Remarkably, the
benchmark MHD models lie close to $\texp/\tcross\sim1$, near the transition
between the impulsive and adiabatic regimes identified in this work.

\item The expansion rate of young stellar systems is not set directly by the SFE
itself, but by the stellar velocity scale already present before gas expulsion.
In our models, the expansion can only draw from the maximum stellar velocity
dispersion reached during the embedded phase, \sigmamax. The quantity $\drdt /
\sigmamax$ is thus better interpreted as a measure of how efficiently the natal
velocity scale is converted into bulk expansion, rather than as a direct tracer
of the natal velocity scale itself. In addition, the comparison with the
benchmark STARFORGE models suggests that the assembly history and phase-space
coherence of the stellar population may also contribute to the later expansion.

\item We examined commonly used observational expansion diagnostics and found
that both the velocity-gradient metric $\kappa$ (when scaled by the true
half-mass radius) and the outward velocity directly trace the
true expansion rate in the ideal case where full kinematic information is
available. In contrast, the ratio $\vout/\sigmavr$ appears to trace how
coherent the expansion is relative to the internal velocity dispersion of the
system. This result still requires calibration against observational biases.
However, it suggests that homogeneous comparisons of expansion strengths between
young stellar systems may already provide useful information about their birth
environments.

\item The expansion rates observed in stellar associations preserve information
about their birth environments. In our models, they represent a lower limit on
the velocity dispersion reached at the onset of gas expulsion, while the
efficiency with which this velocity scale is converted into bulk expansion is
regulated by the SFE and the gas-expulsion timescale, particularly in low-SFE
systems. In nearly fully unbound associations, the present-day expansion rate of
the half-mass radius reaches values of $\sim70\%$ of the embedded velocity
dispersion, decreasing approximately exponentially with gas-expulsion
timescale.

\end{itemize}

These results, however, still need to be tested against observational biases and
more realistic star formation processes including primordial binaries, stellar
evolution and substructure. Such effects may further modulate both the expansion
signatures and the survival of young stellar systems after gas expulsion. The
models presented here provide a baseline to isolate the fundamental dynamical
imprint that embedded star formation environments leave on the observable
kinematics of young stellar populations.

\appendix

\section{Fitting the gas evolution from retention curves}
\label{app:gas_fitting}

We fit the gas model used in this work from the retention of gas inside a set of
fixed initial Lagrangian radii. The goal is to obtain a compact time-dependent
Plummer model, with $M_{\rm pl}(t)$ and $\rpl(t)$, that reproduces the gas
evolution in the region that matters dynamically for the stars.

We define the onset time $t_0$ as the first-star snapshot. At that time, we
measure the radii enclosing a fixed set of gas mass fractions $f_i$. We then
keep these radii, $r_i$, fixed. At later times, we measure how much gas remains
inside those same initial radii. This defines the retention curves
$F_{\rm ret}(f_i,t)$ that we use in the fit.

At each snapshot, we approximate the gas with a Plummer sphere with total mass
$M_{\rm pl}(t)$ and scale radius $\rpl(t)$. We write the mass normalization
relative to onset as
\begin{equation}
S(t)\equiv \frac{M_{\rm pl}(t)}{M_0},
\end{equation}
where $M_0$ is the total gas mass at $t_0$. The fit at each snapshot then
reduces to the two unknowns $\rpl(t)$ and $S(t)$.

For a trial value of $\rpl$, the model predicts a retention fraction at each
fixed shell. At that same $\rpl$, we obtain the best $S$ analytically through a
least-squares fit to the measured retention fractions. This reduces the problem
to a one-dimensional minimization over $\rpl$. Repeating this step at every
snapshot gives the fitted time series $\rpl(t)$ and $S(t)$.

At each snapshot, we fit the retention fractions across all fixed shells to
obtain the Plummer scale radius $\rpl(t)$ and a dimensionless mass factor
$S(t)=M_{\rm pl}(t)/M_0$. We then use the enclosed gas masses from all shells
and all snapshots together in a second, global least-squares step to determine
the single absolute mass scale $M_0$. The fitted Plummer mass series then
follows from
\begin{equation}
M_{\rm pl}(t)=M_0\,S(t).
\end{equation}

The final fitted product is therefore the time series
\begin{equation}
\left\{t,\ M_{\rm pl}(t),\ \rpl(t)\right\}.
\end{equation}
We then use this table either directly as the tabulated gas background or as
the input for the analytic parameterization described in Appendix~A.

We fit the retention curves rather than the full gas density profile because
they directly track how much gas remains inside the initial star-forming
region. This is the part of the gas evolution that enters most directly into
the time-dependent gravitational potential felt by the stars.

\section{Descriptive model for the gas evolution}
\label{app:gas_model}

We model the large-scale gas distribution as a spherically symmetric Plummer
sphere with time-dependent mass $M_{\rm pl}(t)$ and scale radius $\rpl(t)$. We
use this model as a controlled representation of the gas potential. The goal is
not to claim that the gas follows an exact Plummer profile at all times, but to
capture the part of the gas evolution that matters most for the stellar
dynamics: the large-scale and central depth of the background potential.

At each time, the gas density is
\begin{equation}
\rho(r,t)=\frac{3M_{\rm pl}(t)}{4\pi \rpl(t)^3}
\left(1+\frac{r^2}{\rpl(t)^2}\right)^{-5/2},
\end{equation}
with enclosed mass
\begin{equation}
M_{\rm pl}(<r,t)=M_{\rm pl}(t)\,
\frac{r^3}{\left(r^2+\rpl(t)^2\right)^{3/2}}.
\end{equation}

In the workflow used here, this gas model appears in two forms. In the tabulated
version, we fit the MHD gas evolution snapshot by snapshot (see
Appendix~\ref{app:gas_fitting}) and use the resulting time series
\begin{equation}
t,\quad M_{\rm pl}(t),\quad \rpl(t)
\end{equation}
directly into the $N$-body calculation as background potential. In the
parameterized version, we replace this fitted sequence by a compact analytic
model that captures the same broad behaviour with a small number of parameters.

We anchor the analytic model at the onset of star formation, $t_0$, defined as
the first-star snapshot. The parameterized model assumes two phases: an
embedded contraction phase, followed by a post-peak expansion phase. For
$t\leq t_{\rm ge}$, we write
\begin{equation}
\rpl(t)=\rpl{_0}
\sqrt{1-\frac{t-t_0}{t_{\rm col}}},
\end{equation}
where $\rpl{_0}$ is the onset scale radius and $t_{\rm col}$ is the collapse
timescale. For $t>t_{\rm ge}$, we switch to an exponential expansion,
\begin{equation}
\rpl(t)=\rpl(t_{\rm ge})
\exp\left(\frac{t-t_{\rm ge}}{t_{\rm exp}}\right),
\end{equation}
where $t_{\rm exp}$ controls how
rapidly the gas potential opens after the collapse has reached its maximum
concentration.

For the gas mass, we use a simple piecewise function,
\begin{equation}
M_{\rm pl}(t)=M_0+\dot{M}(t-t_0), \qquad t\leq t_{\rm ge},
\end{equation}
and
\begin{equation}
M_{\rm pl}(t)=M_{\rm pl}(t_{\rm ge}), \qquad t>t_{\rm ge}.
\end{equation}
In the parameterized runs used in this work, we take $\dot{M}=0$, so the mass
remains constant and the time dependence enters mainly through $\rpl(t)$.

We determine the parameters of this model from the shell-based retention
analysis of the original MHD simulation. We use the onset properties to fix
$\rpl{_0}$ and $M_0$, and we use the peak of the retention curve at a chosen
calibration shell to define $t_{\rm ge}$ and the maximum retention factor
$F_{\rm ret,max}$. In this work, we take the calibration shell to be the
initial 10\% gas Lagrangian radius, so that $f_{\rm cal}=0.1$ and
$r_{\rm cal}$ is the radius enclosing that fraction at $t_0$. In this
formulation, $F_{\rm ret,max}$ is the quantity used to set the maximum
contraction, while the collapse timescale $t_{\rm col}$ is derived from it. For
$\dot{M}=0$, the value of $F_{\rm ret,max}$ fixes the Plummer radius at the
transition time,
\begin{equation}
\rpl(t_{\rm ge})=
r_{\rm cal}\sqrt{(f_{\rm cal}F_{\rm ret,max})^{-2/3}-1},
\end{equation}
and the collapse timescale then follows from
\begin{equation}
t_{\rm col}=
\frac{t_{\rm ge}-t_0}
{1-\left(\rpl(t_{\rm ge})/\rpl{_0}\right)^2}.
\end{equation}
We then determine the remaining timescale $t_{\rm exp}$ by comparing the
post-peak retention tail of the analytic model to the measured retention curves
and selecting the value that minimizes the residual.

The final parameterized background is therefore specified by the set
\begin{equation}
\{\rpl{_0},\ M_0,\ t_{\rm ge},\ F_{\rm ret,max},\ t_{\rm exp}\},
\end{equation}
with $t_{\rm col}$ derived from these quantities, and with $\dot{M}=0$ in the
runs used here. This provides a compact description of the gas evolution that
preserves the three features that matter most dynamically: the initial depth of
the potential, the amount of contraction before gas expulsion, and the
timescale over which the background potential opens afterward.

\bibliography{bibliography}{}
\bibliographystyle{aasjournalv7}

\end{document}